\documentclass{article}
\usepackage[authoryear]{natbib}
\usepackage{amsthm,amsmath,amssymb,bm}
\usepackage{mathrsfs}
\usepackage{float} 
\usepackage{graphicx}
\usepackage{subfigure}
\usepackage{float}
\usepackage{footnote}
\usepackage{cite}
\usepackage{soul}
\usepackage{authblk}
\usepackage{color,xcolor}
\soulregister{\citet}7
\soulregister{\romannumeral}7
\sethlcolor{yellow}

\newcommand \keywords [1]{\textbf{Keywords}: #1}

\begin{document}
%	\keywords{causal inference; causal effect; multivariate continuous  treatments; entropy balancing.}%\\
	%\noindent \hspace*{-4pc} %{\small\it (Up to five keywords are allowed and should be given in alphabetical order. Please capitalize on the key. }
	%\\
	%\hspace*{-4pc} %{\small\it words)}
	%\\[1pc]
	%\noindent\hspace*{-4.2pc} Supporting Information for this article is available from the author or on the WWW under\break \hspace*{-4pc} \underline{http://dx.doi.org/10.1022/bimj.XXXXXXX} (please delete if not
	%applicable)
	%}%%% semicolon and fullpoint added here for keyword style
	
	\title{Causal Effect Estimation for Multivariate Continuous Treatments}
	\author[1]{Juan Chen}
	\author[1]{Yingchun Zhou \thanks{Corresponding author:  yczhou@stat.ecnu.edu.cn}}
	\affil[1]{Key Laboratory of Advanced Theory and Application in Statistics
				and Data Science-MOE, School of Statistics, East China Normal University.}
	\date{}
	%% Information for the first author.
%	\author[Juan Chen {\it{et al.}}]{Juan Chen\inst{1}} 
	%	\footnote{Corresponding author: {Yingchun Zhou, \sf{e-mail: yczhou@stat.ecnu.edu.cn}}, Phone: +00-999-999-999, Fax: +00-999-999-999}\inst{1}} 
%%	\address[\inst{1}]{Key Laboratory of Advanced Theory and Application in Statistics
%		and Data Science-MOE, School of Statistics, East China Normal University
%		3663 North Zhongshan Road, Shanghai, 200062, P.R. China}
	%%%%    Information for the second author
%	\author[]{Yingchun Zhou\inst{1} \footnote{Corresponding author: {Yingchun Zhou, \sf{e-mail: yczhou@stat.ecnu.edu.cn.}}}}
	%	\address[\inst{2}] {Shanghai Key Laboratory of Maternal Fetal Medicine,Shanghai First Maternity and Infant Hospital, School of Medicine, Tongji University, Shanghai 200092, China}
	%%%%    Information for the third author
	%	\author[]{Jing Hua\inst{2}\footnote{Corresponding author: {Jing Hua, \sf{e-mail: jinghua@tongji.edu.cn.}}}}
	
	%%%%    \dedicatory{This is a dedicatory.}
	%\Receiveddate{zzz} \Reviseddate{zzz} \Accepteddate{zzz} 
		\maketitle  
	\begin{abstract}
		Causal inference is widely used in various fields, such as biology, psychology and economics, etc. In observational studies, we need to balance the covariates before estimating causal effect. This study extends the one-dimensional entropy balancing method to multiple dimensions to balance the covariates. Both parametric and nonparametric methods are proposed to estimate the causal effect of  multivariate continuous treatments and theoretical properties of the two estimations are provided. Furthermore, the simulation results show that the proposed method is better than other methods in various cases. Finally, we apply the method to analyze the impact of the duration and frequency of smoking on medical expenditure. The results show that the frequency of smoking increases medical expenditure significantly while the duration of smoking does not.
	\end{abstract}
	
	%% maketitle must follow the abstract.
                 % Produces the title.

	%% If there is not enough space inside the running head
	%% for all authors including the title you may provide
	%% the leftmark in one of the following three forms:
	
	%% \renewcommand{\leftmark}
	%% {First Author: A Short Title}
	
	%% \renewcommand{\leftmark}
	%% {First Author and Second Author: A Short Title}
	
	%% \renewcommand{\leftmark}
	%% {First Author et al.: A Short Title}
	
	%% \tableofcontents % Produces the table of contents.
	\newpage
	\section{INTRODUCTION}
	
	For decades, causal inference has been widely used in many fields, such as biology, psychology and economics, etc. Most of the current research is based on univariate treatment (binary treatment, multivalued treatment, continuous treatment)  (\citet{imai2014covariate}; \citet{zhu2015boosting}; \citet{fong2018covariate}; \citet{zubizarreta2015stable}; \citet{chan2016globally}; \citet{xiong2017treatment}; \citet{yiu2018covariate}; \citet{dong2021regression}; \citet{hsu2020counterfactual}).  %(Imai and Ratkovic, 2014; Zhu et al., 2015; Zubizarreta, 2015; Chan et al., 2016; Fong et al., 2018; Yiu and Su, 2018; Dong, Lee, and Gou, 2019; Huber, Hsu, Lee, and Lettry, 2020; Colangelo and Lee, 2020). 
	Some research are focused on multivariate categorical treatments, such as factorial designs and conjoint analysis to estimate the main or interaction effect of any combination level of treatments (\citet{2014Causal}; \citet{dasgupta2015causal}). However, sometimes decision-makers are interested in the causal effects of multivariate continuous treatments in real life. For example, when considering the impact of the export and import volume on a country's GDP, one is interested in a bivariate continuous treatment. The methods for multivariate categorical treatments are not suitable to multivariate continuous treatments and there have been few research on this. The goal of this paper is to develop a new method to estimate the causal effect function for multivariate continuous treatments.
	%	That is, the individual is either receiving treatment or not. However, in real life decision makers are also very interested in the causal effects of treatment intensity. For example, when evaluating the impact of a corporate bond purchase plan on market quality, the causal effect at this time depends not only on whether to purchase bonds, but also on the volume of purchases (Boneva, et al., 2018). Therefore, in order to study the causal effect of treatment intensity, it is necessary to expand the binary treatment to multivalue or continuous treatment. In recent years, there has been a lot of research work in this area (Imai and Ratkovic, 2014; Zhu et al., 2015; Zubizarreta , 2015; Chan et al., 2016; Fong et al., 2018; Yiu and Su, 2018; Dong, Lee, and Gou (2019), Huber, Hsu, Lee, and Lettry (2020), Colangelo and Lee (2020 ) ). 
	%	The main interested estimation is the average causal effect, which is the average difference of the response variable to two level of treatment. However, since we can only observed one level of treatment for each individual, we need to take strong ignorability assumption to connect the potential ootcome to the observed values.
	
	\noindent 
	
	A major challenge for inferring the causal effect in observational studies is to balance the confounding covariates, which affect both the treatment and outcome variables. The covariate balancing propensity score method is widely used in controlling for confounding (\citet{rosenbaum1983central}; \citet{rosenbaum1984reducing}; \citet{rosenbaum1985constructing}; \citet{robins2000marginal}; \citet{hirano2004propensity}).
	%Rosenbaum and Rubin, 1983, 1984, 1985; Robins et al., 2000; Hirano et al., 2003). 
	When using the parametric method to model the propensity score, the estimation bias will be large if the model is mis-specified. Therefore, some nonparametric methods for estimating the propensity score have been proposed, such as the kernel density estimation (\citet{robbins2020robust}). In addition, in recent years, some studies have used optimized weighting methods to directly optimize the balance of covariates (\citet{hainmueller2012entropy}; \citet{imai2004causal}; \citet{vegetabile2020optimally}).
	%(Hainmueller 2012; Imai and Ratkovic 2014; Vegetabile et al., 2020). 
	These methods avoid the direct construction of the propensity scores, therefore the obtained estimate achieves higher robustness. One of the methods, the entropy balancing method, has been established as being doubly robust, in that a consistent estimate can still be obtained when one of the two models, either the treatment assignment model or the outcome model, is correctly specified (\citet{zhao2017entropy}).
	%(Zhao and Percival 2017). 
	Furthermore, this method can be easily implemented by solving a convex optimization problem. Therefore, we extend this method to multivariate continuous treatments to balance the covariates in this study.
	
	\noindent
	
	This study has the following three contributions: First, 
	it extends the univariate entropy balancing method to multivariate continuous treatments to balance the covariates. Second, both parametric and nonparametric causal effect estimation methods for multivariate continuous treatments are proposed. Under the parametric framework, a weighted optimization estimation is defined and its theoretical properties are provided. Under the nonparametric framework, B-splines are used to approximate the causal effect function and the convergence rate of the estimation is provided. Third, we apply the proposed method to explore the impact of the duration and frequency of smoking on medical expenditure. The results reveal their relationship, in particular, the frequency of smoking increases medical costs significantly.
	
	\noindent
	
	The remainder of this paper is organized as follows: In Section 2, we introduce the motivating example in this study. In Section 3, the entropy balancing for multivariate treatment (EBMT) method is proposed. In Section 4, theoretical properties of the parametric and nonparametric estimation are shown. In section 5, methods for variance estimation and confidence interval construction are provided. In section 6, a numerical simulation is performed to evaluate the properties of the EBMT method. In Section 7, the EBMT method is applied to the real data analysis. The conclusions and discussions are summarized in Section 8.

	%	\noindent Names of software packages and website addresses should be written in {\tt{Courier new, i.e. Stata, the R package
	%MASS, http://www.biometrical-journal.com.}}
	\section{MOTIVATING EXAMPLE}
	In this section, we introduce an observational study that contains two-dimensional treatments that motivates our methodology. The causal relationship between smoking and medical expenditure has long been a hot topic of research. Most studies are confined to a binary treatment (smoker/non-smoker), and estimate its effect on medical costs (\citet{rubin2000statistical}; \citet{larsen1999analysis}; \citet{2000Statistical} ). However, the binary treatment is too rough in describing the status of a smoker. A better way is to regard the duration and frequency of smoking as bivariate continuous treatments and study their causal effect on medical costs, which motivates this study.
	
	The data we used is extracted from the 1987 National Medical Expenditure Survey (NMES), which is originally studied by \citet{Elizabeth2003Disease}.  The origional study does not directly estimate the causal effect of smoking on medical expenditure, instead, it first estimates the effect of smoking on certain diseases and then examines the increase of medical costs due to the diseases. \citet{Elizabeth2003Disease} proposed a variable, called $packyear$, to measure the cumulative smoking that combines the duration and frequency of smoking, which is defined as
	\begin{equation*}
	packyear = \frac{\text{number of cigrattes per day}}{20} \times \text{number of years smoked.}
	\end{equation*}  
	\citet{2004Causal} directly estimates the causal effect of smoking on medical expenditures with a univariate continuous treatment $packyear$ and with a bivariate continuous treatment (the duration and frequency of smoking), respectively. However, when considering the bivariate treatment, the authors estimate their causal effects on medical costs separately, which may lose important information. 
		\noindent
		
		Motivated by this example, we develop a method to estimate the causal effect jointly when the treatment is multivariate.

	\section{ENTROPY BALANCING FOR MULTIVARIATE TREATMENTS}
	In this section, we introduce the EBMT method to obtain weights that can balance covariates and  both parametric and nonparametric approaches are developed to estimate the causal effect function.
	\subsection{Notation and assumptions}
	Suppose that the treatment for subject $i$ is $\textbf{T}_i$, whose support is $\mathcal{T} \subset \mathcal{R}^p$. $\textbf{X}_i \in \mathcal{R}^q$ denotes the observed covariates, where $p,q$ denote the dimensions of the treatments and covariates, respectively. Suppose that for each subject, there exists a potential outcome $Y_i(\textbf{t})$ for all $\textbf{t} \in \mathcal{T}$. The observed outcome is defined as $Y_i = Y_i(\textbf{t})$ if $\textbf{T}_i = \textbf{t}$. Assume that a sample of observations $\lbrace Y_i, \textbf{T}_i, \textbf{X}_i \rbrace$ for $i \in \lbrace 1,....,n \rbrace$ is independently drawn from a joint distribution $f(Y,\textbf{T}, \textbf{X})$. For notational convenience, the treatments and covariates are assumed to be standardized.
	
	\noindent 
	
	To perform causal inference with the observational data, the following three standard assumptions are made (\citet{hirano2004propensity}; \citet{imai2004causal}).\\
	%(Hirano and Imbens, 2004; Imai and van Dyk, 2004):
	
	\noindent
	\textbf{Assumption 1 (Strong Ignorability)}:\\
	\\
	$\textbf{T}_i \perp Y_i(\textbf{t}) \mid \textbf{X}_i$, which means that the treatment assignment is independent of the counterfactual outcomes, conditional on covariates. This implies that there is no unmeasured confounding.
	\\
	
	\noindent
	\textbf{Assumption 2 (Positivity)}:\\
	\\
	$f_{\textbf{T} \mid \textbf{X}}(\textbf{T}_i = \textbf{t} \mid \textbf{X}_i ) > 0$ for all $\textbf{t} \in \mathcal{T}$, and the conditional density $f(\textbf{T}_i \mid \textbf{X}_i )$ is called the generalized propensity score (\citet{imbens2000role}).\\
	%(Imbens, 2000).
	
	\noindent
	\textbf{Assumption 3 (SUVTA)}:\\
	\\
	Assume that there is no interference among the units, which means that each individual's outcome depends only on their own level of treatment intensity.
	\\
	
	The goal of this paper is to estimate the causal effect function $\mathbb{E}(Y(\mathbf{t}))$ for a multivariate continuous treatment $\textbf{T}$ based on the above assumptions. In order to estimate the causal effect function with observational data, we first define the stabilized weight as\\
	\begin{equation*}
		 w_i = \frac{f(\textbf{T}_i)}{f(\textbf{T}_i \mid \textbf{X}_i)},
	\end{equation*}
	then under the strong ignorability assumption, one can estimate the causal effect function based on the stabilized weight by using parametric or nonparametric method.
	
	\subsection{Entropy balancing for multivariate treatments}
	The entropy balancing method (\citet{hainmueller2012entropy})
	%(Hainmueller 2012) 
	is used to determine the optimal weight for inferring causal effects. It has been used for univariate treatment and here we extend this method to multivariate treatments. 
	
	\noindent
	
	Note that the stabilized weight \\
	\begin{equation}
		w_i = \frac{f(\textbf{T}_i)}{f(\textbf{T}_i \mid \textbf{X}_i)},
	\end{equation}
	satisfies the following conditions:
	\begin{equation}
		\mathbb{E}(w_i) = \iint \frac{f(\textbf{T}_i)}{f(\textbf{T}_i \mid \textbf{X}_i)}f(\textbf{T}_i, \textbf{X}_i)d\textbf{T}_id\textbf{X}_i = 1,
	\end{equation}
	\begin{equation}
		\begin{split}
			\mathbb{E}(w_i\textbf{T}_i\textbf{X}_i^{'}) &= \iint \frac{f(\textbf{T}_i)}{f(\textbf{T}_i \mid \textbf{X}_i)}\textbf{T}_i\textbf{X}_i^{'}f(\textbf{T}_i, \textbf{X}_i)d\textbf{T}_id\textbf{X}_i \\
			&=\int \lbrace \frac{f(\textbf{T}_i)}{f(\textbf{T}_i \mid \textbf{X}_i)}\textbf{T}_if(\textbf{T}_i \mid \textbf{X}_i)d\textbf{T}_i \rbrace \textbf{X}_i^{'}f(\textbf{X}_i)d\textbf{X}_i\\
			&=\mathbb{E}(\textbf{T}_i)\mathbb{E}(\textbf{X}_i^{'}) = \mathbf{0}.
		\end{split}
	\end{equation}
	Similarly, it can be shown that weighting with $w_i$ also preserves the marginal means of $\textbf{T}_i$ and $\textbf{X}_i$, which means
	\begin{equation}
		\mathbb{E}(w_i\textbf{T}_i)=\mathbb{E}(\textbf{T}_i)=\mathbf{0},
	\end{equation}
	\begin{equation}
		\mathbb{E}(w_i\textbf{X}_i)=\mathbb{E}(\textbf{X}_i)=\mathbf{0}.
	\end{equation}
	Hence we can set the sample conditions as
	\begin{equation}
		\sum_{i=1}^{n}w_ig(\textbf{T}_i,  \textbf{X}_i) = \mathbf{0}, \ \sum_{i=1}^{n}w_i = 1,  \ w_i > 0  \  \  \ \ \forall i = 1,...,n,
	\end{equation}
	where $g(\textbf{T}_i,  \textbf{X}_i)  = [\text{vec}((\textbf{T}_i\textbf{X}_i^{'}))^{'},\textbf{T}_i,  \textbf{X}_i]^{'}$, whose dimensions are $pq+p+q$. Actually, this framework can be generalized to multivariate treatments with categorical variables or combinations of categorical and continuous variables, which is beyond the scope of this article.

	\noindent
	
	We now choose weights that satisfy the conditions in Equation (6), while minimizing the Kullback-Leibler divergence:
	\begin{gather}
		\text{min}_w \sum_{i=1}^{n}w_ilog(\frac{w_i}{v_i})  \notag                                   
	\end{gather}
	s.t.
	\begin{gather}
		\sum_{i=1}^{n}w_ig(\textbf{T}_i,  \textbf{X}_i) = \mathbf{0}, \	\sum_{i=1}^{n}w_i = 1,  \ w_i > 0  \ \forall i = 1,..., n.  
	\end{gather}
	
	In Equation (7), $v_i$ denotes the base weights, which is equal to $\frac{1}{n}$ in this study.
	
	\noindent
	
	We adopt a method which is similar to the standard Lagrange multiplier technique for solving this optimization problem. Using this method, we obtain the weighting function in terms of the Lagrange multipliers $\gamma$ as 
	\begin{equation}
		w_i = \frac{v_iexp(-\gamma^{'}g(\textbf{T}_i,  \textbf{X}_i))}{\sum_{i=1}^{n}v_iexp(-\gamma^{'}g(\textbf{T}_i,  \textbf{X}_i))}.
	\end{equation}
	The proof can be found in Appendix A.2. Based on these weights, a new objective function is obtaind,
	\begin{equation}
		\text{min}_\gamma \ log(\sum_{i=1}^{n}v_iexp(-\gamma^{'}g(\textbf{T}_i,  \textbf{X}_i))).
	\end{equation}
	This new dual objective function can be optimized by using an efficient convex optimization algorithm. Note that when the dimension of covariates or treatments is high, the numerical algorithm might fail to find a solution. Hence, this method is most suitable to low-dimensional treatments and covariates. High-dimensional covariates or treatments case will be considered in future work.
	
	\subsection{CAUSAL EFFECT ESTIMATION}
	In this subsection, both parametric and nonparametric approaches are developed to estimate the causal effect function. A weighted optimization estimation is defined under the parametric framework and B-splines are used to approximate the causal effect function under the nonparametric framework.
	\noindent
	
	\subsubsection{Parametric method}
	The causal effect function is parametrized as $s(\textbf{t};\theta)$, we assume that it has a unique solution  $\theta^* \in \mathcal{R}^J$ (with $J \in \mathbb{N}$) defined as
	\begin{equation}
		\theta^* = \text{agrmin}_\theta \int_{\mathcal{T}} \mathbb{E}[Y(\textbf{t})-s(\textbf{t};\theta)]^2f_\textbf{T}(\textbf{t})d\textbf{t}.
	\end{equation}
	The difficulty for solving Equation (10) is that the potential outcome $Y(\textbf{t})$ is not observed for all $\textbf{t}$. Hence, Proposition 1 is proposed to connect the potential outcome with the observed outcome. \\
	
	\noindent
	\textbf{Proposition 1} \\  \\
	Under Assumption 1, it can be shown that
	\begin{equation}
		\mathbb{E}[w(Y-s(\textbf{T};\theta))^2] = \int_\mathcal{T}  \mathbb{E}[Y(\textbf{t})-s(\textbf{t};\theta)]^2f_\textbf{T}(\textbf{t})d\textbf{t}.
	\end{equation}
	The proof of Proposition 1 can be found in Appendix A.1. Note that $Y(\textbf{t})$ on the right hand side of Equation (11) represents the potential outcome and $Y$ on the left hand side represents the observed outcome. Proposition 1 indicates that by inserting $w$ on the left hand side of Equation (11), one can represent the objective function with the potential outcome by that with the observed outcome. Therefore, the true value $\theta^*$ is also a solution for the weighted optimization problem:
	\begin{equation}
		\theta^* = \text{argmin}_\theta  \mathbb{E}[w(Y-s(\textbf{T};\theta))^2].
	\end{equation}
	This result implies that the true value $\theta^*$ can be identified from the observational data.
	One can obtain the estimator based on the sample, which is
	\begin{equation}
		\hat{\theta} = \text{argmin}_\theta \sum_{i=1}^{n} \hat{w_i}(Y_i-s(\textbf{T}_i; \theta))^2.
	\end{equation}
	
	%Similarly, although we can also get fully nonparametric estimation of $s(\mathbf{t})$ within this framework, we do not consider it in this article and it will be considered in a future work. Next, we will introduce another nonparametric estimation of $s(\mathbf{t})$ .
	
	\subsubsection{Nonparametric method}
	Suppose $\mathbb{E}(Y(\textbf{t})) = s(\textbf{t})$. In a similar manner to the proof of Proposition 1, it can be shown that 
	\begin{equation*}
		\mathbb{E}(wY \mid \textbf{T}=\textbf{t}) = \mathbb{E}(Y(\textbf{t})).
	\end{equation*}
	\noindent
	
	In this paper, B-splines are used to approximate $s(\textbf{t})$. For notational convenience, assume that without loss of generality, all B-splines of order $r$ are defined on an extended partition associated with a uniform partition of $m$ knots. Following \citet{schumaker2007spline}, %Schumaker (2007), 
	we denote the B-spline basis functions on the $i$th component of $\textbf{T}$ as $B_j(\textbf{t}_k) (j=1,...,M; k=1,...,p; M=m+r)$. Furthermore, define
	\begin{equation}
		B_{k_1,...,k_p}(\textbf{t}) = \prod_{j=1}^{p} B_{k_j}(\textbf{t}_j), \forall \ 1 \leq k_1....,k_p\leq M.
	\end{equation}
	Let $B(\textbf{t})$ be the $Q\equiv M^p$-dimensional vector consisting of all product functions of the form (14) and $Z_n = (B(\textbf{T}_1),...,B(\textbf{T}_n))^{'}$, then the B-spline estimation of $s(\textbf{t})$ is determined by 
	\begin{equation}
		\hat{s}(\textbf{t}) = B(\textbf{t})^{'}\hat{\beta}, 
	\end{equation}
	where
	\begin{equation}
		\hat{\beta} = (Z_n^{'}Z_n)^{-1}Z_n^{'}\hat{W}Y \notag, \ \hat{W}= diag(\hat{w}_1,....., \hat{w}_n).
	\end{equation}

	\section{LARGE SAMPLE PROPERTIES}
	To establish the large sample properties of the proposed estimators in section 3.3, 
	%according to the theorm 3 of Ai,et al.(2020), we can first conclude that the estimated weights function $\hat{w}$ is consistent and we can derive its convengence rate. Furthermore, 
	we show the consistency and limiting distribution of the optimization estimator $\hat{\theta}$ under the parametric framework and the convergence rate of the nonparametric estimator $\hat{s}(\textbf{t})$. The following assumptions are made: \\

	\noindent
	\textbf{Assumption 4}
	\renewcommand \theenumi{\roman{enumi}}
	\renewcommand \labelenumi{(\theenumi)}
	\begin{enumerate}
		\item The support $\mathcal{T}$ of $\textbf{T}$ is a compact subset of $\mathcal{R}^p$ and the support $\mathcal{X}$ of $\textbf{X}$ is a compact subset of $\mathcal{R}^q$.
		\item There exist two positive constants $a_1$ and $a_2$, such that 
		\\
		$ 0< a_1 \leq w \leq a_2 < \infty , \forall (\textbf{t},\textbf{x}) \in \mathcal{T} \times \mathcal{X} $.
	\end{enumerate}
	
	%	\noindent
	%	\textbf{Assumption 5} \\ \\
	%There exist $\gamma_{(pq+p+q)} \in \mathcal{R}^{pq+p+q}$ such that 
	%$\text{sup}_{(\textbf{t},\textbf{x})} \mid e^{-w}-\gamma^{'}g(\textbf{t},\textbf{x}) \mid = O(1)$. \\
	%There exist $M_{(p+1)(q+1)} \in \mathcal{R}^{(p+1)(q+1)}$ and a positive constant $\alpha > 0$, such that \\
	%$sup_{(\textbf{t},\textbf{x})} \mid g^{'-1}(w)-h(\textbf{t})M_{(p+1)(q+1)}e(\textbf{x}) \mid = O((p+1)(q+1)^{-\alpha})$

	\noindent
	\textbf{Assumption 5}
	\begin{enumerate}
		\item The parameter space $\Theta$ is a compact set and the true parameter $\theta_0$ is in the interior of $\Theta$.
		\item $(Y-s(\textbf{t};\theta))^2$ is continuous in $\theta$ , $\mathbb{E}[\text{sup}_\theta(Y-s(\textbf{t};\theta))^2] < \infty $ and $\text{sup}_\theta \mathbb{E}[(Y-s(\textbf{t};\theta))^4]  < \infty $.
	\end{enumerate}

	\noindent 
	\textbf{Assumption \ 6}
	\begin{enumerate}
		\item $s(\mathbf{t};\theta)$ is twice continuously differentiable in $\theta \in \Theta$ and let $h(\mathbf{t};\theta) \equiv \bigtriangledown_\theta s(\mathbf{t};\theta)$.
		\item $\mathbb{E} \lbrace w(Y-s(\mathbf{T};\theta))h(\mathbf{T};\theta) \rbrace $ is differentiable with respect to $\theta$ and \\
			$U \equiv - \bigtriangledown_\theta \mathbb{E} \lbrace w(Y-s(\mathbf{T};\theta))h(\mathbf{T};\theta) \rbrace \mid_{\theta=\theta^*}$ is nonsingular.
		\item $\mathbb{E}[\text{sup}_\theta \mid Y-s(\textbf{t};\theta) \mid^{2+\delta}] < \infty $ for some $\delta >0$ and there extists some finite positive constants $a$ and $b$ such that $\mathbb{E}[\text{sup}_{\theta_1:  \mid \mid \theta_1-\theta \mid \mid  < \delta_1} \mid s(\textbf{t};\theta_1) - s(\textbf{t};\theta) \mid^{2}]^{1/2} < a\cdot \delta_1^b $ for any $\theta \in \Theta$ and any small $\delta_1 >0$.
	\end{enumerate}

	\noindent
	
	Assumption 4(\romannumeral1) restricts the treatments and covariates within a bounded range. This is convenient for calculating the convergence rate under the $L_\infty$ norm. Assumption 4(\romannumeral2) restricts the weight function to be bounded away from zero. Given Assumption 4(\romannumeral1), Assumption 4(\romannumeral2) is equivalent to the positivity assumption. %Assumption 5 restricts the convergence rate of the approximation error, it is established that the convergence rate of the weight function is bounded by the convergence rate of the error (\citet{ai2021unified}).
	%(Ai et al.,(2020)). 
	Assumption 5(\romannumeral1) is a commonly used assumption in nonparametric regression. Assumption 5(\romannumeral2) is an envelope condition applicable to the uniform law of large numbers. Assumption 6 (\romannumeral1) and (\romannumeral2) impose sufficient regularity conditions on the causal effect function and its derivative function. Assumption 6 (\romannumeral3) is a stochastic equicontinuity condition, which is needed for establishing weak convergence (\citet{1994Chapter}). Based on these assumptions, the following theorems are established.\\
	
	\noindent 
	\textbf{Theorem 1}\\ 
	\begin{enumerate}
		\item Under Assumptions 1-5, $\mid\mid \hat{\theta}-\theta^* \mid\mid \ \to_p 0$. 
		\item Under Assumptions 1-6, $\sqrt{n}(\hat{\theta}-\theta^*) \to_d N(0,V)$, where 
		\begin{equation*}
		V = 4U^{-1}\cdot \mathbb{E} \lbrace w^2(Y-s(\mathbf{T};\theta^*))^2h(\mathbf{T};\theta^*)h(\mathbf{T};\theta^*)^{'} \rbrace\cdot U^{-1} 
		\end{equation*}
	\end{enumerate}

	\noindent
	Under the nonparametric framework, Theorem 2 is established to obtain the convergence rate of the estimate $\hat{s}(\textbf{t})=B(\textbf{t})^{'}\hat{\beta}$.\\
	
	\noindent
	\textbf{Theorem \ 2} \  Suppose $\text{sup}_{\textbf{t} \in \mathcal{T}} \mid s(\textbf{t}) - B(\textbf{t})^{'}\beta^*) \mid = O(Q^{-\tilde{\alpha}})	$  holds for some $\tilde{\alpha}$ and $\beta^* \in \mathbb{R}^Q$, and Assumptions 1-4 hold, then 
	\begin{equation*}
		\begin{split}
			&\int \mid \hat{s}(\textbf{t}) - s(\textbf{t}) \mid^2f_\textbf{T}(\textbf{t})d\textbf{t} = O_p(Q^{-2\tilde{\alpha}}+\frac{Q}{n}) \\
			&\text{sup}_\textbf{t} \mid \hat{s}(\textbf{t}) - s(\textbf{t}) \mid =O_p( Q^{-\tilde{\alpha}}+\sqrt{\frac{Q}{n}}) 
		\end{split}
	\end{equation*}
	\noindent
	%	where $b_Q$ satisfies that $\mid\mid \text{sup}_\textbf{t}B(\textbf{t}) \mid\mid \leq b_Q$.\\
	
	\noindent
	The proofs of Theorem 1 and Theorem 2 can be found in Appendix A.3.
	
	\section{VARIANCE ESTIMATION AND CONFIDENCE INTERVAL }
	\subsection{Variance estimation}
	To conduct statiatical inference, a consistent estimator of the covariance matrix $V$ is needed, which can be obtained by replacing $w$ and $\theta^*$ by their consistent estimators. Specifically, one can write $U$ as
	\begin{equation*}
		\begin{split}
		U&=- \bigtriangledown_\theta \mathbb{E} \lbrace w(Y-s(\mathbf{T};\theta))h(\mathbf{T};\theta) \rbrace \mid_{\theta=\theta^*}\\
			&= -\mathbb{E}\lbrace w \bigtriangledown_\theta[(Y-s(\mathbf{T};\theta))h(\mathbf{T};\theta)]\rbrace \mid_{\theta=\theta^*} \\
			&=- \mathbb{E}\lbrace w [-h(\mathbf{T};\theta)h(\mathbf{T};\theta)^{'}+(Y-s(\mathbf{T};\theta))\bigtriangledown_\theta h(\mathbf{T};\theta)] \rbrace\mid_{\theta=\theta^*} \\
			&= \mathbb{E}\lbrace w [h(\mathbf{T};\theta^*)h(\mathbf{T};\theta^*)^{'}-(Y-s(\mathbf{T};\theta^*))\bigtriangledown_\theta h(\mathbf{T};\theta^*)] \rbrace.
		\end{split}
	\end{equation*}
	Hence, 
	\begin{equation*}
	\hat{U} = \frac{1}{n} \sum_{i=1}^{n}\hat{w}_i [h(\mathbf{T}_i;\hat{\theta}) h(\mathbf{T}_i;\hat{\theta})^{'} - (Y_i-s(\mathbf{T}_i;\hat{\theta})) \bigtriangledown_\theta h(\mathbf{T}_i;\hat{\theta})].
	\end{equation*}
	Then we have
	\begin{equation*}
	\hat{V} = 4 \hat{U}^{-1}  \cdot \lbrace \frac{1}{n}\sum_{i=1}^{n} \hat{w}_i^2(Y_i-s(\mathbf{T}_i;\hat{\theta}))^2 h(\mathbf{T}_i;\hat{\theta}) h(\mathbf{T}_i;\hat{\theta})^{'} \rbrace \cdot \hat{U}^{-1}.
	\end{equation*}
	According to Theorem 1(\romannumeral1) and Theorem 3 in \citet{ai2021unified}, we have $\mid\mid \hat{\theta} - \theta^* \mid\mid \to_p 0 $ and \\ $\text{sup} \mid \hat{w}-w \mid = o_p(1)$, which implies that $\hat{V}$ is consistent.
	
	\subsection{Confidence interval}
According to the asymptotic normality of $\hat{\theta} = (\hat{\theta}_1,\dots, \hat{\theta}_J)^{'}$, one can construct the 95$\%$ confidence interval of $\hat{\theta}_j$ as 
	\begin{equation}
		[\hat{\theta}_j - 1.96\cdot \hat{SE}_j, \hat{\theta}_j + 1.96\cdot \hat{SE}_j ],\  \forall j = 1,\dots ,J,
	\end{equation}
where $\hat{SE}_j = \hat{V}_{jj}/\sqrt{n}$ is the standard error of $  \hat{\theta}_j $, $ \hat{V}_{jj}$ is the $(j,j)$-element of $\hat{V}$.

	\noindent
	
	Alternatively, one can also construct the confidence interval using bootstrap method. Suppose the $b^{\text{th}}$ bootstrap sample $\lbrace Y_i^{(b)}, \mathbf{T}_i^{b}, \mathbf{X}_i^{(b)} \rbrace, i= 1,\dots ,n; b= 1,\dots ,B$ is sampled with replacement from the origional sample $\lbrace Y_i, \mathbf{T}_i, \mathbf{X}_i\rbrace, i=1,\dots,n$ with uniform distribution. Denote $\hat{\theta}^{(b)}$ as the estimator of $\theta^*$ based on the $b^{\text{th}}$ bootstrap sample. Arrange the $B$ bootstrap estimators in order from smallest to largest,  then the 95$\%$ confidence interval can be defined as
	\begin{equation}
		[\hat{\theta}_j^{(0.025B)}, \hat{\theta}_j^{(0.975B)}],  \ \forall j = 1,\dots ,J.
	\end{equation}
	The difference of the two methods is that the first method (16) relies on the asymptotic normality result while the second method (17) does not.

	\section{SIMULATION}
	To examine the properties of the proposed estimators under finite samples, simulation studies are performed under different data settings. The main motivation in designing the simulation settings is to compare the proposed method with three other methods when the treatment assignment model and the outcome model are linear and nonlinear in various ways.
	
	\subsection{Assessment criteria of covariate balance and effect estimation }
	
	\noindent
	Assume that the treatments follow the multiple multivariate linear regression model:
	\begin{equation}
		\mathbf{T}_i= B^{'}\mathbf{X}_i+\epsilon_i, i=1,....,n,
	\end{equation}
	where $\mathbf{T}_i$  and $\mathbf{X}_i$ denote the $p$-dimensional treatments and $q$-dimensional covariates, respectively.  $B_{q\times p} = (\bm{\beta}_1 :\bm{\beta}_2:...:\bm{\beta}_q)$ represents the coefficient matrix, and $\epsilon_i \sim N_p(\mathbf{0},\Sigma)$. The MLEs of $B$  are $\hat{B}_{q\times p} = (\mathbf{X}^{'}\mathbf{X})^{-1}\mathbf{X}^{'}\mathbf{T}$, where $\mathbf{T} = (\mathbf{T}_1^{'},....., \mathbf{T}_n^{'})^{'}, \ \mathbf{X} = (\mathbf{X}_1^{'},....., \mathbf{X}_n^{'})^{'}$.
	
	\noindent
	
	To assess the performance of the covariate balance, the following hypothesis test is considered:
	\begin{equation*}
		H_0 :\bm{ \beta}_1 = \bm{\beta}_2 = ..... = \bm{\beta}_q = \mathbf{0}. 
	\end{equation*}
	The likelihood ratio statistic is: 
	\begin{equation*}
		\Lambda = \frac{\mid SSE \mid}{\mid SSE+SSH \mid},
	\end{equation*}
	where $SSE= \mathbf{T}^T\mathbf{T}-\hat{B}^T\mathbf{X}^T\mathbf{T}$ and $SSH = \hat{B}^T\mathbf{X}^T\mathbf{T} - n\bar{\mathbf{T}}\bar{\mathbf{T}}^T$ denote the  the residual sum of squares and the predicted sum of squares, respectively. 
	
	\noindent
	When $n$ is large, 
	\begin{equation*}
		-2log(\Lambda) \to \chi_{qp}^2.
	\end{equation*}
	Hence, it is harder to reject the null hypothesis when $-2log(\Lambda)$ is closer to zero, which implies that the covariate balance performs well.
	
	\noindent
	
	Next, we focus on the outcome model, which is linear and nonlinear in the treatments, respectively.
	\begin{gather}
		Y_i=r(\mathbf{X}_i)+\mathbf{T}_i\beta+\varepsilon_i, \\
		Y_i= r(\mathbf{X}_i) + s(\mathbf{T}_i) +\varepsilon_i,
	\end{gather}
	\noindent
	where $\varepsilon_i \sim N(0,\sigma^2), i=1,...,n$.
	
	\noindent
	
	For the linear outcome model, the bias and RMSE of the coefficients are used to assess the performance of the different methods in various cases .
	For the nonparametric outcome model, we use
	%\begin{gather*}
	%RMSE_{linear} = \sqrt{\frac{1}{p}\sum_{i=1}^{p}(\hat{\beta_i}-\beta_i)}
	%\end{gather*}
	%and \\
	\begin{gather*}
		\text{RMSE}= \sqrt{\frac{1}{n}\sum_{i=1}^{n}(\hat{s}(\mathbf{t})_i-s(\mathbf{t})_i)^2}
	\end{gather*}
	\noindent
	as the assessment criteria.
	
	\subsection{Data generating process}
	
	The data generating process is described in this subsection. For the treatment assignment model, two major cases are considered, which are T-2d-L, T-2d-NL, respectively, where "T" stands for treatment, "2d" stands for two-dimension, "L" and "NL" stands for linear and nonlinear in the covariates. For the outcome model, linear and nonlinear cases are studied, which are considered for each treatment assignment model. For notational convenience, "Specification $YK$" is used to denote the $K\text{th}$ outcome model.
	
	\noindent
	
	Since the covariates are shared across all scenarios, their data generating process is first described. Specifically, we independently draw 5 covariates from a multivariate normal distribution with mean 0, variance 1 and covariance 0.2, that is,
	
	\[  \mathbf{X} = (X_1,....,X_5)^{'} \sim  N_5(\mu,\Sigma) \ \text{with}\ \mu= \begin{pmatrix} 0\\  \vdots \\0 \end{pmatrix} \text{and}\ \Sigma=\begin{pmatrix} 1&0.2 &\dots& 0.2 \\0.2&1 &0.2 \dots & 0.2 \\  \hdotsfor{4} \\0.2&0.2&\dots &1 \end{pmatrix}_{5\times5.} \]
	\vspace{12pt}
	
In the first simulation setting, assume that both the treatment assignment model "T-2d-L” and the outcome model are linear in the covariates, with true data generating process given by
	
	\noindent
	\begin{equation}
	\textbf{T-2d-L}: \mathbf{T}_i= B_1^{'}\mathbf{X}_i+\mathbf{\epsilon}_i,
	\end{equation}
	Specification $Y1$:
	\begin{gather*}
		Y_i =T_{i1}+T_{i2}+X_{i1}+0.1X_{i2}+0.1X_{i5}+\varepsilon_i.
	\end{gather*}
	%\vspace{12pt}
	%\noindent
	Specification $Y2$:
	\begin{gather*}
		Y_i =T_{i1}+T_{i2}+0.2T_{i1}T_{i2}+X_{i1}+0.1X_{i2}+0.1X_{i5}+\varepsilon_i.
	\end{gather*}
	Specification $Y3$:
	\begin{gather*}
		Y_i =T_{i1}+T_{i2}+(T_{i1}-T_{i2})^2+X_{i1}+0.1X_{i2}+0.1X_{i5}+\varepsilon_i,
	\end{gather*}
	where 
	\begin{gather*}
	\mathbf{T}_i = (T_{i1},T_{i2})^{'},\ 
	B_1=\begin{bmatrix}
				1 & 1\\
				0 & 0.2 \\
				0.2&0 \\
				0&0 \\
				0&0
			\end{bmatrix}_{5 \times 2}, \\varepsilon_i \sim N(0,2^2),\\
	\mathbf{\epsilon}_i \sim  N_2(\mu,\Sigma) \ 
	\text{with} \ \mu= \begin{pmatrix} 0\\0 \end{pmatrix}  \text{and} \ \Sigma=\begin{pmatrix} 3&0.8 \\0.8&3 \end{pmatrix} .
	\end{gather*}	
	\noindent
	
	In the second simulation setting, consider the case when the treatment assignment model is misspecified and the outcome model remains the same as in the first simulation setting, that is, the treatment model is nonlinear in the covariates, denoted as "T-2d-NL":
	
	\noindent
	\begin{equation}
		\textbf{T-2d-NL}: \mathbf{T}_i= B_1^{'}\mathbf{X}_i+B_2^{'}(\mathbf{X}_i*\mathbf{X}_i)+\epsilon_i,
	\end{equation}
	\noindent
where $B_1$ and $\mathbf{\epsilon}_i$ are the same as in Equation (22), $\textbf{X}*\textbf{X}$ represents the Hadamard product of matrices \textbf{X} and \textbf{X}, and its corresponding element is $(\textbf{X}*\textbf{X})_{ij} = (x_{ij}x_{ij})$, $B_2= \begin{bmatrix}
			1 & 1\\
			0&0\\
			\vdots&  \vdots \\
			0 & 0
		\end{bmatrix}_{5 \times 2}$.

	In the third simulation setting, we consider misspecification of the outcome model by including nonlinear covariate term in the outcome model. In this case, the generating process for the treatment model remains the same as in the first simulation setting. Specifically, the misspecified outcome model is defined as
	
	\noindent
Specification$Y4$:
	\begin{gather*}
		Y_i =T_{i1}+T_{i2}+(X_{i1}+1)^3+0.1X_{i2}+0.1X_{i5}+\varepsilon_i.
	\end{gather*}
	\noindent
Specification $Y5$:
	\begin{gather*}
	Y_i =T_{i1}+T_{i2}+0.2T_{i1}T_{i2}+(X_{i1}+1)^3+0.1X_{i2}+0.1X_{i5}+\varepsilon_i.
	\end{gather*}
	\noindent
	Specification $Y6$:
	\begin{gather*}
	Y_i =T_{i1}+T_{i2}+(T_{i1}-T_{i2})^2+(X_{i1}+1)^3+0.1X_{i2}+0.1X_{i5}+\varepsilon_i,
	\end{gather*}
	\noindent
	where $\varepsilon_i \sim N(0,2^2)$.
	
	The last simulation setting considers the misspecification of both the treatment assignment model and the outcome model, which are defined in the second simulation setting and the third simulation setting, respectively.

	\subsection{Simulation results}
	In this subsection, simulation results are compared between the proposed method (EBMT), the regression covariate adjustment method (RCAM), entropy balancing for univariate treatment method (EBUT) and the multivariate generalized propensity score method (mvGPS) (\citet{2020Causal}), where RCAM refers to estimating the causal effect function by adjusting for all covariates in the outcome model, EBUT refers to handling each single treatment seperately, mvGPS refers to the method that estimates the causal effect function assuming that the treatments follow a multivariate normal distribution. 
	
	Specifically, the performance of covariate balancing are compared among EBMT, mvGPS and Unweighted methods, since RCAM does not balance covariates and EBUT does not fit multiple multivariate regression so that the balancing effect cannot be measured. When there is no interaction effect bewteen treatments, the estimation accuracy of all four methods is compared. But when there is interaction between treatments, only three methods (EBMT, RCAT, mvGPS) are compared since the method EBUT can not deal with this case. In addition, the 95$\%$ coverage probability and the average width of the confidence interval are considered. These are obtained using the bootstrap method with $B=500$ iterations based on Equation (17) for each effect estimation.  For each numerical setting, 1000 independent simulation experiments are run.  
		
		Figure 1 shows the results for the covariate balancing of the linear two-dimensional treatment assignment model (top) and nonlinear two-dimensional treatment assignment model (bottom) under different settings. The statistics $-2log(\Lambda)$ produced by the EBMT method are almost zero, while those of the mvGPS and Unweighted methods are far away from zero. These indicate that the EBMT method balances the covariates well and is robust when the treatment assignment model is linear and nonlinear in the covariates.
	
	Table 1 shows the results of the causal effect estimation for the linear outcome model without interaction between the two-dimensional treatments under different simulation settings. It can be seen that when both the treatment assignment model and outcome model are correctly specified, the mean bias of the EBMT method is the smallest among the four methods. The RMSE of the EBMT method is the second smallest while that of RCAT is the smallest. The RMSE of all methods decreases as the sample size increases. Among the four methods mvGPS performs the worst. When one of the two models is misspecified, the mean bias of EBMT is the smallest in both cases, and the RMSE of EBMT is the smallest when the treatment model is misspecified. When both the two models are misspecified, all methods fail.
		
		Table 2 shows the results of the causal effect estimation for the linear outcome model with interaction between the treatments. As can be seen, the results for the main effect are similar to those in Table 1, that is, EBMT performs the best in most cases. For the interaction effect, the mean biases are slightly larger in most cases. Similarly, all methods fail when both the two models are misspecified. %Indeed, when the sample size reaches 10,000, the mean bias of the interaction effect estimated by both methods can reach 0.001.
	\begin{figure}[h]
		\centering
		\subfigure{
			\begin{minipage}{12cm}
				\centering
				\includegraphics[width=10cm]{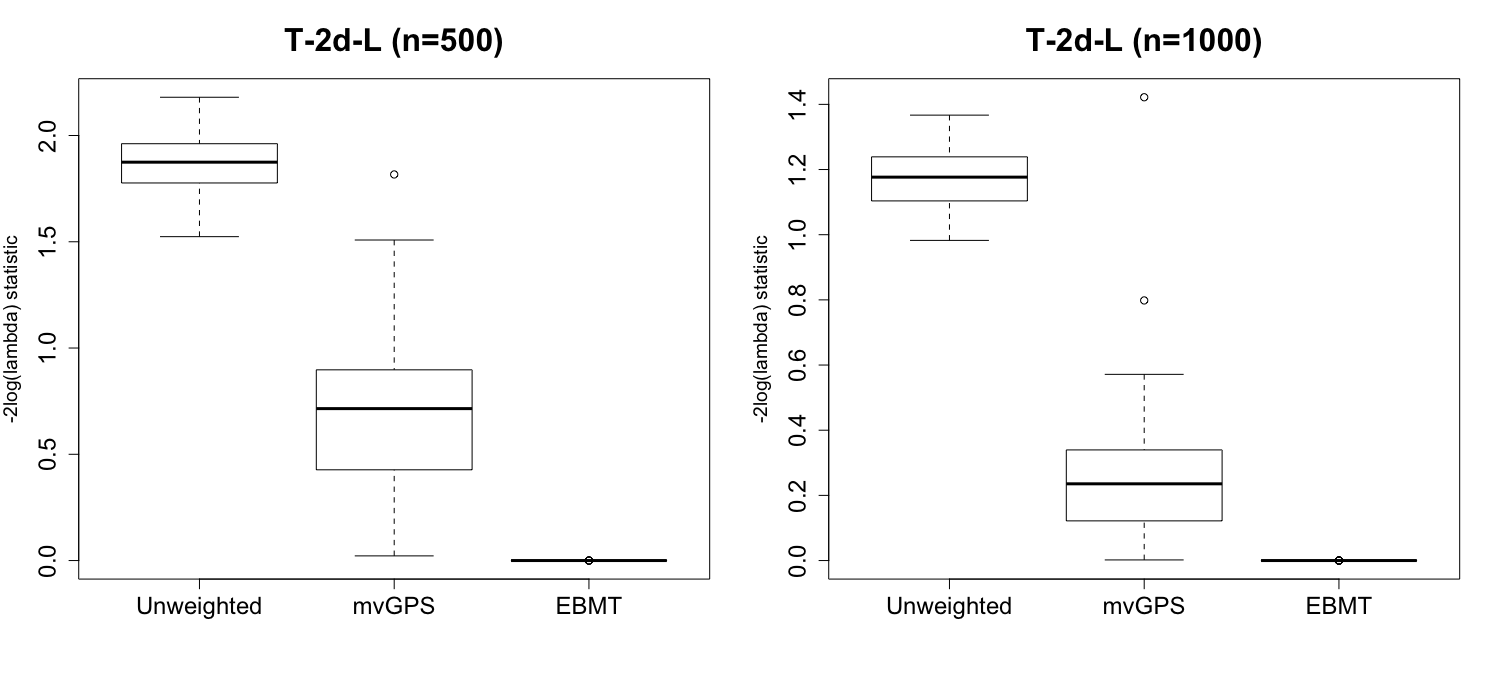} 
			\end{minipage}
		}
		\subfigure{
			\begin{minipage}{12cm}
				\centering
				\includegraphics[width=10cm]{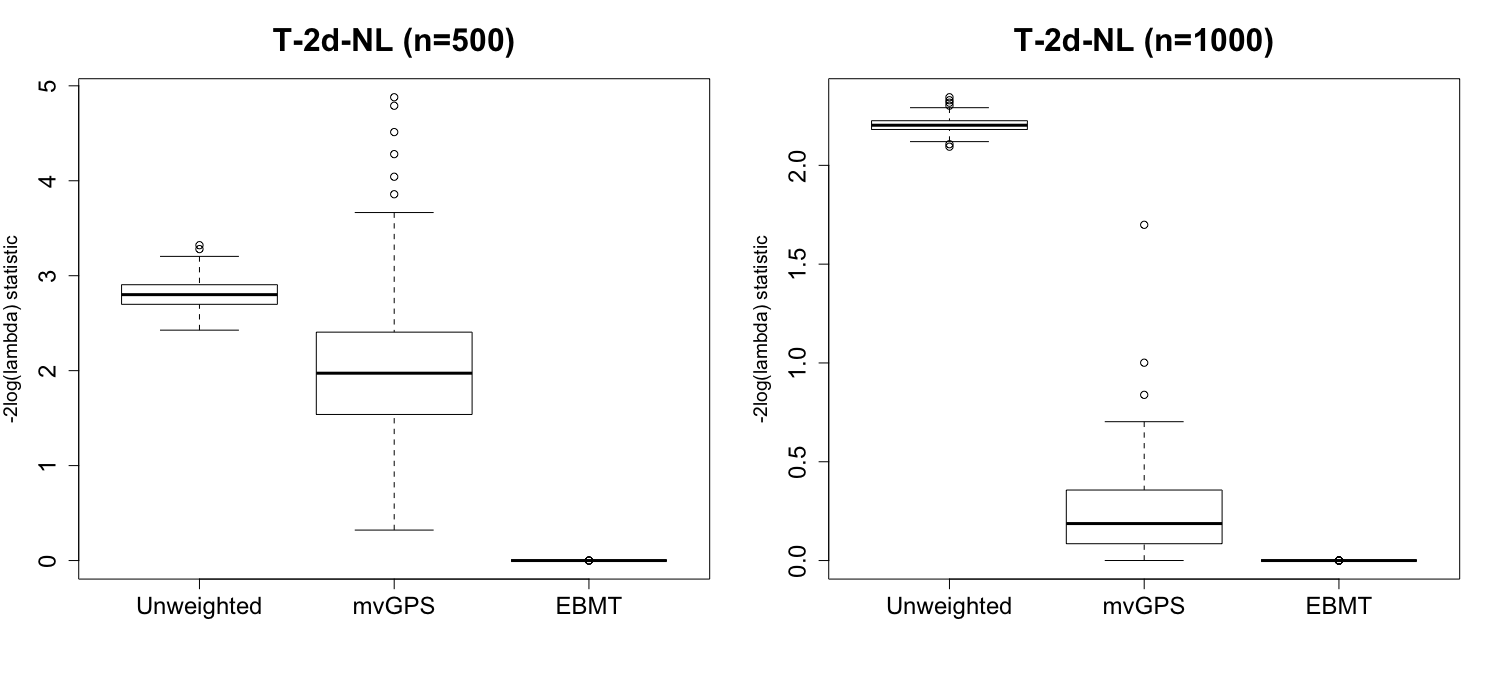} 
			\end{minipage}
		}
		\caption{Performance comparison of covariate balancing for two-dimensional treatments.}
	\end{figure}

	\begin{table}[h]
		\begin{center}
			\caption{Mean bias and RMSE of the coefficients in the linear outcome model without interaction between the two-dimensional treatments.} 
			\renewcommand \arraystretch{1.5}
			\begin{tabular}{ccccc}
				\hline
				\textbf{Method} & \multicolumn{4}{c}{\textbf{Both \ $E(T\mid X)$ \ and \ $E(Y\mid X)$ \ correctly \ specified}} \\
				\hline
				& \multicolumn{2}{c}{\textbf{$\hat{\beta}_1$ ($\beta_1^* = 1$) }}& \multicolumn{2}{c}{ \textbf{$\hat{\beta}_2$ ($\beta_2^* = 1$) }}\\ \hline
				&$n=500$ & $n=1000$ &$n=500$ & $n=1000$\\ \hline
				& Bias/RMSE & Bias/RMSE & Bias/RMSE & Bias/RMSE \\ \hline
			EBMT&-0.002/0.086&0.001/0.052& 0.005/0.188 &0.004/0.119 \\ \hline
			RCAM&-0.005/0.058&-0.004/0.043& 0.011/0.104& 0.010/0.081\\ \hline
				EBUT&-0.003/0.084& -0.002/0.058& 0.008/0.206 & 0.007/0.133 \\ \hline
			mvGPS&-0.012/0.132&0.006/0.082& 0.151/0.319 &0.035/0.171 \\ \hline
				&\multicolumn{4}{c}{  \textbf{$E(T\mid X)$ \ correctly \ specified, \ $E(Y \mid X)$ \ misspecified}} \\ \hline
				& \multicolumn{2}{c}{\textbf{$\hat{\beta}_1$ ($\beta_1^* = 1$) }}& \multicolumn{2}{c}{ \textbf{$\hat{\beta}_2$ ($\beta_2^* = 1$) }}\\ \hline
				&$n=500$ &$n=1000$&$n=500$& $n=1000$\\ \hline
				&Bias/RMSE& Bias/RMSE& Bias/RMSE&Bias/RMSE \\ \hline
				EBMT& -0.008/0.125 &-0.001/0.087&0.019/0.276 & 0.028/0.178 \\ \hline
				RCAM&-0.011/0.151& 0.007/0.105&0.031/0.333& -0.047/0.194\\ \hline
				EBUT&0.014/0.189& 0.022/0.144&0.023/0.284& 0.034/0.180\\ \hline
				mvGPS& -0.075/0.767& -0.016/0.591& 1.035/1.408& 0.762/1.051\\ \hline
				&\multicolumn{4}{c}{\textbf{$E(T\mid X)$ \ misspecified, \ $E(Y \mid X)$ \ correctly \  specified}}\\ \hline
				& \multicolumn{2}{c}{\textbf{$\hat{\beta}_1$ ($\beta_1^* = 1$) }}& \multicolumn{2}{c}{ \textbf{$\hat{\beta}_2$ ($\beta_2^* = 0.8$) }}\\ \hline
				&$n=500$&$n=1000$& $n=500$ &$n=1000$\\ \hline
				& Bias/RMSE &Bias/RMSE& Bias/RMSE&Bias/RMSE\\ \hline
				EBMT& -0.001/0.063 &0.003/0.048&0.005/0.118& -0.004/0.074  \\ \hline
				RCAM&-0.003/0.054& 0.007/0.043& 0.009/0.079& -0.005/0.058 \\ \hline
				EBUT&-0.013/0.105& 0.003/0.067& 0.008/0.137& -0.007/0.108  \\ \hline
				mvGPS&-0.063/0.616&-0.015/0.321& -0.369/0.795& -0.178/0.482 \\ \hline
				&\multicolumn{4}{c}{\textbf{Both \ $E(T\mid X)$ \  and \ $E(Y\mid X)$ \ misspecified }} \\ \hline
				& \multicolumn{2}{c}{\textbf{$\hat{\beta}_1$ ($\beta_1^* = 1$) }}& \multicolumn{2}{c}{ \textbf{$\hat{\beta}_2$ ($\beta_2^* = 0.8$) }}\\ \hline
				&$n=500$ &$n=1000$ & $n=500$&$n=1000$\\ \hline
				& Bias/RMSE &Bias/RMSE& Bias/RMSE & Bias/RMSE\\ \hline
				EBMT& 0.006/0.119 &0.009/0.090&1.032/1.058 & 0.965/0.994 \\ \hline
				RCAM&0.001/0.116& 0.006/0.085& 2.144/2.174& 2.139/2.154 \\ \hline
				EBUT&0.008/0.147& 0.020/0.100& 1.082/1.105& 1.080/1.091  \\ \hline
				mvGPS& -0.076/0.814& 0.228/1.762&-1.359/1.742& -1.937/2.650 \\ \hline
			\end{tabular}
		\end{center}
		%	\footnotesize{Note: $\hat{\beta}_1$ and $\hat{\beta}_2$ denote the estimated coefficients of $T_1$ and $T_2$, respectively.} \\
	\end{table}

	\begin{table}[h]
		\begin{center}
			\caption{Mean bias and RMSE of the coefficients in the linear outcome model with interaction between the two-dimensional treatments. }
			\renewcommand \arraystretch{1.5}
			\begin{tabular}{ccccccc}
				\hline
				\textbf{Method} & \multicolumn{6}{c}{\textbf{Both \ $E(T\mid X)$ \ and \  $E(Y\mid X)$ \ correctly \ specified}} \\
				\hline
				& \multicolumn{2}{c}{\textbf{$\hat{\beta}_1$ ($\beta_1^* = 1$) }}& \multicolumn{2}{c}{ \textbf{$\hat{\beta}_2$ ($\beta_2^* = 0.8$) }}& \multicolumn{2}{c}{ \textbf{$\hat{\beta}_{12}$ ($\beta_{12}^* = 0.2$) }}\\ \hline
				&$n=500$ & $n=1000$& $n=500$ & $n=1000$&$n=500$ &$ n=1000$\\ \hline
				& Bias/RMSE & Bias/RMSE & Bias/RMSE & Bias/RMSE& Bias/RMSE& Bias/RMSE\\ \hline
				EBMT& 0.002/0.093 &-0.004/0.064&0.003/0.184 & 0.002/0.137& -0.013/0.086 & -0.011/0.062  \\ \hline
			RCAM&0.009/0.067&-0.007/0.43& 0.006/0.113&0.009/0.081 & -0.006/0.036& -0.005/0.022  \\ \hline
			mvGPS&0.011/0.139& -0.017/0.123&0.153/0.317& 0.086/0.271 & -0.008/0.076& -0.007/0.068  \\ \hline
				& \multicolumn{6}{c}{\textbf{$E(T\mid X)$ \ correctly \ specified, \ $E(Y \mid X)$ \ misspecified}} \\ \hline
				& \multicolumn{2}{c}{\textbf{$\hat{\beta}_1$ ($\beta_1^* = 1$) }}& \multicolumn{2}{c}{ \textbf{$\hat{\beta}_2$ ($\beta_2^* = 0.8$) }}& \multicolumn{2}{c}{ \textbf{$\hat{\beta}_{12}$ ($\beta_{12}^* = 0.2$) }}\\ \hline
				&$n=500$ & $n=1000$ & $n=500$ & $n=1000$&$n=500$ &$ n=1000$ \\ \hline
				& Bias/RMSE & Bias/RMSE & Bias/RMSE & Bias/RMSE& Bias/RMSE & Bias/RMSE \\ \hline
				EBMT& 0.008/0.111 &-0.007/0.085&-0.005/0.283 & -0.017/0.194 & -0.056/0.170 & -0.031/0.143  \\ \hline
				RCAM&0.024/0.151& 0.012/0.106& -0.044/0.306& 0.027/0.228 & -0.589/0.604& -0.606/0.620 \\ \hline
			mvGPS&0.176/0.796&-0.067/0.430& 0.575/0.431&0.621/0.387 & -0.066/0.258& -0.004/0.193  \\ \hline
				& \multicolumn{6}{c}{\textbf{$E(T\mid X)$ \ misspecified, \ $E(Y \mid X)$ \ correctly \  specified}} \\ \hline
				& \multicolumn{2}{c}{\textbf{$\hat{\beta}_1$ ($\beta_1^* = 1$) }}& \multicolumn{2}{c}{ \textbf{$\hat{\beta}_2$ ($\beta_2^* = 0.8$) }}& \multicolumn{2}{c}{ \textbf{$\hat{\beta}_{12}$ ($\beta_{12}^* = 0.2$) }}\\ \hline
				&$n=500$ & $n=1000$ & $n=500$ & $n=1000$&$n=500$ &$ n=1000$ \\ \hline
				& Bias/RMSE & Bias/RMSE & Bias/RMSE & Bias/RMSE& Bias/RMSE & Bias/RMSE \\ \hline
				EBMT&0.003/0.022& 0.002/0.016 &-0.005/0.037&-0.003/0.025 &-0.004/0.009& -0.002/0.008 \\ \hline
			RCAM&0.010/0.030 &0.009/0.028& -0.012/0.025&-0.008/0.018& -0.002/0.003& -0.001/0.004  \\ 	\hline
				mvGPS&0.002/0.043& 0.006/0.032& 0.109/0.128& -0.093/0.100& -0.009/0.014& -0.007/0.012  \\ \hline
				& \multicolumn{6}{c}{\textbf{Both \ $E(T\mid X)$ \ and \ $E(Y\mid X)$ \ misspecified}} \\ \hline
				& \multicolumn{2}{c}{\textbf{$\hat{\beta}_1$ ($\beta_1^* = 1$) }}& \multicolumn{2}{c}{\textbf{$\hat{\beta}_2$ ($\beta_2^* = 0.8$) }}& \multicolumn{2}{c}{ \textbf{$\hat{\beta}_{12}$ ($\beta_{12}^* = 0.2$) }}\\ \hline
				&$n=500$ & $n=1000$ & $n=500$ & $n=1000$&$n=500$ &$ n=1000$ \\ \hline
				& Bias/RMSE & Bias/RMSE & Bias/RMSE & Bias/RMSE& Bias/RMSE & Bias/RMSE\\ \hline
			EBMT& 0.093/0.177&0.153/0.207&1.435/1.507 & 1.617/1.673 &-0.104/0.186& -0.156/0.198 \\ \hline
				RCAM&-0.263/0.278& -0.254/0.266& 0.951/0.967& 0.995/1.004& 0.266/0.269& 0.254/0.257  \\ \hline
				mvGPS&0.417/0.947&0.571/0.994&-0.888/1.807& -0.989/1.474 & -0.146/0.304& -0.168/0.245 \\ \hline
			\end{tabular}
		\end{center} 
	\end{table}

	%boostrap interval
	\begin{table}[h]
		\begin{center}
			\caption{Confidence interval estimation of the coefficients in the linear outcome model without interaction between the two-dimensional treatments . }
			\renewcommand \arraystretch{1.5}
			\begin{tabular}{ccccc}
				\hline
				\textbf{Method} & \multicolumn{4}{c}{\textbf{Both \ $E(T\mid X)$ \ and \ $E(Y\mid X)$ \ correctly \ specified}} \\
				\hline
				& \multicolumn{2}{c}{\textbf{$\hat{\beta}_1$ ($\beta_1^* = 1$) }}& \multicolumn{2}{c}{ \textbf{$\hat{\beta}_2$ ($\beta_2^* = 1$) }}\\ \hline
				&$n=500$ & $n=1000$ & $n=500$ & $n=1000$\\ \hline
				&95 $\%$ CP/AW &95 $\%$ CP/AW &95 $\%$ CP/AW & 95 $\%$ CP/AW\\ \hline
			EBMT& 0.951/0.325 &0.956/0.222& 0.946/0.681 & 0.953/0.478 \\ \hline
			RCAM&0.926/0.222& 0.942/0.162&0.934/0.438& 0.952/0.318\\ \hline
				EBUT& 0.944/0.0.306& 0.953/0.226& 0.877/0.734 & 0.914/0.527 \\ \hline
		mvGPS& 0.948/0.434& 0.954/0.340&0.758/0.723 & 0.812/0.598 \\ \hline
				&\multicolumn{4}{c}{\textbf{$E(T\mid X)$ \ correctly \ specified, \ $E(Y \mid X)$ \ misspecified}} \\ \hline
				& \multicolumn{2}{c}{\textbf{$\hat{\beta}_1$ ($\beta_1^* = 1$) }}& \multicolumn{2}{c}{ \textbf{$\hat{\beta}_2$ ($\beta_2^* = 1$) }}\\ \hline
				&$n=500$ & $n=1000$ & $n=500$ & $n=1000$\\ \hline
				&95 $\%$ CP/AW &95 $\%$  CP/AW & 95 $\%$ CP/AW & 95 $\%$ CP/AW\\ \hline
			EBMT&0.943/0.429 &0.952/0.290&0.933/0.870 & 0.942/0.644 \\ \hline
				RCAM&0.914/0.630& 0.936/0.414& 0.925/1.307& 0.941/0.819\\ \hline
				EBUT&0.933/0.606& 0.941/0.418& 0.904/0.942& 0.917/0.690\\ \hline
				mvGPS& 0.887/1.478& 0.912/1.087& 0.527/2.243 & 0.634/1.822 \\ \hline
				&\multicolumn{4}{c}{\textbf{$E(T\mid X)$ \ misspecified, \ $E(Y \mid X)$ \ correctly \ specified}} \\ \hline
				& \multicolumn{2}{c}{\textbf{$\hat{\beta}_1$ ($\beta_1^* = 1$) }}& \multicolumn{2}{c}{ \textbf{$\hat{\beta}_2$ ($\beta_2^* = 0.8$) }}\\ \hline
				&$n=500$ & $n=1000$ &$n=500$ & $n=1000$\\ \hline
				& 95 $\%$ CP/AW & 95 $\%$ CP/AW & 95 $\%$ CP/AW &95 $\%$ CP/AW\\ \hline
				EBMT& 0.943/0.263 &0.953/0.189&0.948/0.381 & 0.952/0.284  \\ \hline
				RCAM&0.923/0.221& 0.945/0.163& 0.927/0.299& 0.941/0.218 \\ \hline
			EBUT&0.945/0.339& 0.956/0.240& 0.914/0.523& 0.923/0.381  \\ 	\hline
				mvGPS& 0.932/1.240& 0.935/1.232& 0.847/1.568 & 0.866/1.389 \\ \hline
				&\multicolumn{4}{c}{\textbf{Both \ $E(T\mid X)$ \ and \ $E(Y\mid X)$ \ misspecified}} \\ \hline
				& \multicolumn{2}{c}{\textbf{$\hat{\beta}_1$ ($\beta_1^* = 1$) }}& \multicolumn{2}{c}{ \textbf{$\hat{\beta}_2$ ($\beta_2^* = 0.8$) }}\\ \hline
				&$n=500$& $n=1000$ &$n=500$ & $n=1000$\\ \hline
				& 95 $\%$ CP/AW & 95 $\%$ CP/AW & 95 $\%$ CP/AW & 95 $\%$ CP/AW\\ \hline
				EBMT& 0.914/0.379 &0.943/0.278&0.029/0.717 & 0.034/0.526  \\ \hline
				RCAM&0.961/0.436& 0.965/0.299& 0.001/1.169& 0.001/0.772  \\ \hline
				EBUT&0.886/0.454& 0.911/0.334& 0.012/0.726& 0.020/0.532  \\ \hline
				mvGPS& 0.771/2.254& 0.892/2.074& 0.042/3.293 & 0.038/2.683 \\ \hline
			\end{tabular}
		\end{center}
		%	\footnotesize{Note: $\hat{\beta}_1$ and $\hat{\beta}_2$ denote the estimated coefficients of $T_1$ and $T_2$, respectively.} \\
	\end{table}
	
	\begin{table}[h]
		\begin{center}
			\caption{Mean RMSE comparison of the nonparametric outcome model of the two-dimensional treatments. }
			\renewcommand \arraystretch{1.5}
			\begin{tabular}{ccccc}
				\hline
			\textbf{Method} & \multicolumn{2}{c}{\textbf{Both \ $E(T\mid X)$ \ and \ $E(Y\mid X)$ \ correctly \ specified}} \\
				\hline
				&$n=500$ & $n=1000$ \\ \hline
			EBMT& 0.933 &0.828 \\ \hline
				mvGPS& 1.630& 1.298 \\ \hline
				&\multicolumn{2}{c}{\textbf{$E(T\mid X)$  \ misspecified,  \ $E(Y \mid X)$  \ correctly \ specified}} \\ \hline
				&$n=500$ & $n=1000$ \\ \hline
			EBMT&3.219&2.937 \\ \hline
			mvGPS& 4.802& 4.480 \\ \hline
				&\multicolumn{2}{c}{\textbf{$E(T\mid X)$ \ misspecified,  \ $E(Y \mid X)$ \ correctly \ specified}} \\ \hline
				&$n=500$ & $n=1000$ \\ \hline
			EBMT& 1.258&0.968  \\ \hline
				mvGPS&1.788& 2.784 \\ \hline
				&\multicolumn{2}{c}{\textbf{Both \ $E(T\mid X)$  \ and  \ $E(Y\mid X)$  \ misspecified}} \\ \hline
				&$n=500$ & $n=1000$ \\ \hline
			EBMT& 6.043&5.703 \\ \hline
				mvGPS& 7.286&7.537 \\ \hline
			\end{tabular}
		\end{center}
		%	\footnotesize{Note: $\hat{\beta}_1$ and $\hat{\beta}_2$ denote the estimated coefficients of $T_1$ and $T_2$, respectively.} \\
	\end{table}
	
	\noindent
	
	Table 3 shows the results of the 95$\%$ coverage probability and average width of the bootstrap confidence interval of the causal effect estimation. It can be seen that the 95$\%$ coverage probability of the EBMT method is nearly identical to 0.95 and it's larger than the other methods under different scenarios except for the case that both the treatment assignment model and outcome model are misspecified. For all methods, the average width of the confidence interval shrinks as the sample size grows. 
		
		\noindent
		
		Table 4 shows the mean RMSE of the estimated causal effect function for the nonparametric outcome model under different scenarios. Since only the EBMT and mvGPS methods can be applied to such problem, the results of those two methods are compared. It can be seen that EBMT method enjoys a better performance compared with mvGPS method when both the treatment assignment model and the outcome model are correctly specified or either one of the two models is correctly specified. Both methods fail when both the two models are misspecificed.

	The simulation results can be summarized as follows. (\romannumeral1) 
	In the case that both the treatment assignment model and outcome model are correctly specified, EBMT has the smallest mean bias of the main effect and slightly larger RMSE than RCAT. (\romannumeral2) In the case that the treatment assignment model is correctly specified and the outcome model is misspecified, EBMT has the smallest mean bias and RMSE of all effects. (\romannumeral3) In the case that both the treatment assignment model and outcome model are misspecified, all methods fail. (\romannumeral4) In the case of nonparametric outcome model, EBMT has the smallest mean RMSE.

	\section{APPLICATION}
	We now apply the EBMT method to the motivating example described in Section 2. The goal is to analyze the impact of the duration and frequency of smoking on medical expenditures.
		\noindent
		
		The pre-treatment covariates $\mathbf{X} $ includes age at the time of the survey, age when the individual started smoking, gender, race, marital status, education level, census region, poverty status and seat belt usage, which are the covariates used in Imai et al. (2004). The treatments we are interested in are the duration and frequency of smoking, and the outcome is log(Total Medical Expenditure). Similar to the setting of Imai et al. (2004), we assume a linear outcome model and conduct a complete-case analysis by discarding all missing data, yielding a sample of 7847 smokers.  
		
		Before estimating the causal effect, we first examine the covariate balancing of EBMT and mvGPS methods by conducting a multiple multivariate regression. The statistic $-2log(\Lambda) $ defined in section 6.1 is $3.95e^{-8}$ for EBMT method, 2.27 for mvGPS method and 2.31 for Unweighted method,  which shows that EBMT balances covariates very well while mvGPS does not balance well.
	\noindent
	
	The causal effect function is estimated by the parametric method assuming a linear model of the outcome. Bootstrap method is used to obtain the standard error and confidence interval for the parameter estimates. Each of the 1000 bootstrap replicates estimates weight using EBMT, mvGPS and EBUT methods, since RACM does not estimate weight. Then the outcome model is estimated by the four methods EBMT, RACM,EBUT and mvGPS. 
	%	\begin{figure}[h]
	%	\centering
	%	\centering
	%	\includegraphics[width=10cm]{/Users/yilimideyangguang/Desktop/多维连续处理/A-C-B.png} 
	%	\caption{\hl{Performance comparison of covariates balancing for bootstrap replicates.}}
	%	\end{figure}
	
	\noindent
	
	Table 5 shows the estimated causal effect of the duration and frequency of smoking on medical expenditure as well as its standard error and confidence interval. All methods indicate that the duration of smoking has no significant impact on medical expenditure since their confidence intervals all cover zero while most methods agree that the frequency of smoking increases medical expenditure significantly except mvGPS. Besides, the width of confidence interval of the estimated causal effect based on EBMT method is smallest. Compared with univariate analysis using a single variable $packyear$, the analysis of the bivariate treatments provides more information that the significant effect of $packyear$ attributes mostly to the frequency of smoking rather than to its duration. 
	
	\begin{table}[h]
		\begin{center}
			\caption{Causal effect estimation of the duration and frequency of smoking on meadical expenditure. }
			\renewcommand \arraystretch{1.5}
			\begin{tabular}{ccccccc}
				\hline
				\textbf{Method} & \multicolumn{3}{c}{\textbf{Duration}}& \multicolumn{3}{c}{ \textbf{Frequency }}\\ \hline
				&Estimate& SE& 95$\%$ CI& Estimate & SE& 95$\%$ CI \\ \hline
				EBMT& 0.003&0.002& (-0.003,0.004)& 0.007&0.002&(0.003,0.009)\\ \hline
				RCAM&-0.001& 0.004& (-0.007,0.007)& 0.006&0.003&(0.001,0.011)\\ \hline
				EBUT& 0.002& 0.004&(-0.009,0.006)& 0.007&0.002&(0.004,0.011) \\ \hline
				mvGPS& -0.073&0.169&(-0.079,0.104) & 0.018&0.047&(-0.079,0.104) \\ \hline
				
			\end{tabular}
		\end{center}
		%	\footnotesize{Note: $\hat{\beta}_1$ and $\hat{\beta}_2$ denote the estimated coefficients of $T_1$ and $T_2$, respectively.} \\
	\end{table}

	\section{CONCLUSION AND DISCUSSION}
	In this study, we extend the one-dimensional entropy balancing method to multidimensional treatments. In addition, parametric and nonparametric methods are developed to estimate the causal effect and their theoretical properties are provided. The simulation results show that the proposed method balances covariates well and produces a smaller mean bias compared with other methods. In the real data analysis, the EBMT method is applied to investigate causal relationship between the duration and frequency of smoking on medical expenditure, which indicates that the frequency of smoking inceases medical costs significantly.  
	
	In this paper, we mainly consider the causal effect function $\mathbb{E}(Y(t))$ as the estimand for continuous treatment as it is general. The average treatment effect or average partial effect can also be considered, but they can be easily estimated based on the estimates of causal effect function $\mathbb{E}(Y(t+\triangle t)-Y(t))$,  $\frac{\mathbb{E}Y(t+\triangle t)-\mathbb{E}Y(t)}{\triangle t}$ (\citet{2008A}). Indeed, the causal effect function provides a complete description of the causal effect, rather than a summary measure.

	\noindent
	
	Future works include: (1) Estimating causal effect in the case of high-dimensional covariates or treatments; (2) Considering treatments involves other types of complex data, such as longitudinal data and functional data; (3) Extending this method to the multidimensional outcome variables.
	
	\section*{ACKNOWLEDGEMENT}
	The work was supported by National Natural Science Foundation of China (project number: 11771146, 11831008), the 111 Project (B14019) and Program of Shanghai Subject Chief Scientist (14XD1401600). 
	The work was supported by National Natural Science Foundation of China (project number: 11771146, 11831008), the 111 Project (B14019) and Program of Shanghai Subject Chief Scientist (14XD1401600).

	\newpage
	\nocite{*}
	\bibliographystyle{apalike}
	\bibliography{ref}
	
	\newpage
	\section*{APPENDIX }
	
	\subsection*{A.1.\enspace Proof of Proposition 1}
	Using the law of total expectation and Assumption 1, we can deduce that
	\begin{gather*}
		\begin{split}
			\mathbb{E}[w(Y-s(\textbf{T};\theta))^2]&=E[\frac{f(\textbf{T})}{f(\textbf{T}\mid\textbf{X})}(Y-s(\textbf{T};\theta))^2]\\
			&= \mathbb{E}(\mathbb{E}[\frac{f(\textbf{T})}{f(\textbf{T}\mid\textbf{X})}(Y-s(\textbf{T};\theta))^2] \mid \textbf{T}=\textbf{t},\textbf{X}=\textbf{x} )\\
			&= \mathbb{E}(\frac{f(\textbf{t})}{f(\textbf{t}\mid\textbf{x})}\mathbb{E}([(Y-s(\textbf{T};\theta))^2] \mid \textbf{T}=\textbf{t},\textbf{X}=\textbf{x}) )\\
			&= \int_{\mathcal{T}\times \mathcal{X}}\frac{f(\textbf{t})}{f(\textbf{t}\mid \textbf{x})}\mathbb{E}[(Y(\textbf{T})-s(\textbf{T};\theta))^2 \mid \textbf{T} = \textbf{t}, \textbf{X}= \textbf{x}]f(\textbf{t}\mid \textbf{x})d\textbf{t}d\textbf{x}\\
			&=\int_{\mathcal{T}\times \mathcal{X}}\mathbb{E}[(Y(\textbf{T})-s(\textbf{T};\theta))^2 \mid \textbf{T} = \textbf{t}, \textbf{X}= \textbf{x}]f(\textbf{t})f(\textbf{x})d\textbf{t}d\textbf{x}\\
			&= \int_{\mathcal{T}\times \mathcal{X}}\mathbb{E}[(Y(\textbf{t})-s(\textbf{t};\theta))^2 \mid \textbf{X}= \textbf{x}]f(\textbf{t})f(\textbf{x})d\textbf{t}d\textbf{x}\\
			&= \int_{\mathcal{T}\times \mathcal{X}}\mathbb{E}[(Y(\textbf{T})-s(\textbf{T};\theta))^2 f(\textbf{t})d\textbf{t} \quad (\text{using Assumption 1}).
		\end{split}
	\end{gather*}
	Hence, we complete the proof of Proposition 1.
	
	\subsection*{A.2.\enspace Derivation of optimization procedure}
	Considering the optimization problem defined in Section 3.2:
	\begin{gather}
		\text{min}_w \sum_{i=1}^{n}w_ilog(\frac{w_i}{v_i}) \qquad \notag 
	\end{gather}
	s.t.
	\begin{gather}
		\sum_{i=1}^{n}w_ig(\textbf{T}_i,  \textbf{X}_i) = \mathbf{0}, \	\sum_{i=1}^{n}w_i = 1,  \ w_i > 0  \ \forall i.                                            
	\end{gather}
	Using the standard Lagrange multiplier method to solve this optimization problem, we  construct the Lagrangian as 
	\begin{equation}
		\mathcal{L}(w_i,\alpha,\beta) = \sum_{i=1}^{n}w_i log(\frac{w_i}{v_i})+(\alpha-1)(\sum_{i=1}^{n}w_i-1)+\gamma^{'}\sum_{i=1}^{n}w_ig(\textbf{T}_i,  \textbf{X}_i),
	\end{equation}
	where $\alpha$ and $\beta$ are Langrage multipliers.
	Taking derivatives with respect to $w_i$, the first order condition is
	\begin{gather*}
		\begin{split}
			\frac{\partial}{\partial w_i}\mathcal{L}(w_i,\alpha,\beta) &= (log(\frac{w_i}{v_i})+1)+\alpha-1+\gamma^{'}g(\textbf{T}_i,  \textbf{X}_i)\\
			&= log(\frac{w_i}{v_i})+\alpha+\gamma^{'}g(\textbf{T}_i,  \textbf{X}_i)\\
			&=0.
		\end{split}
	\end{gather*}
	Then we have
	\begin{gather*}
		w_i \propto v_iexp(-[\gamma^{'}g(\textbf{T}_i,  \textbf{X}_i)+\alpha]).
	\end{gather*}
	Enforcing the constraint that the sum of weights must be one, it follows that
	\begin{gather}
		\begin{split}
			w_i&= \frac{ v_iexp(-[\gamma^{'}g(\textbf{T}_i,  \textbf{X}_i)+\alpha])}{\sum_{i=1}^{n} v_iexp(-[\gamma^{'}g(\textbf{T}_i,  \textbf{X}_i)+\alpha])}\\
			&=\frac{ v_iexp(-[\gamma^{'}g(\textbf{T}_i,  \textbf{X}_i)])}{\sum_{i=1}^{n} v_iexp(-[\gamma^{'}g(\textbf{T}_i,  \textbf{X}_i)])}.
		\end{split}
	\end{gather}
	Next, for notational simplicity, we redefine the weights as a parameter $\lambda_i$ divided by a normalizing constant $\lambda = \sum_{i=1}^{n}\lambda_i$, that is,
	\begin{gather*}
		w_i = \frac{ v_iexp(-[\gamma^{'}g(\textbf{T}_i,  \textbf{X}_i)])}{\sum_{i=1}^{n} v_iexp(-[\gamma^{'}g(\textbf{T}_i,  \textbf{X}_i)])} \equiv \frac{\lambda_i}{\lambda}.
	\end{gather*}
	Taking the log transformation, we obtain that 
	\begin{gather*}
		log(\lambda_i) = log(v_i)-\gamma^{'}g(\textbf{T}_i,  \textbf{X}_i),
	\end{gather*}
	implying
	\begin{gather*}
		\gamma^{'}g(\textbf{T}_i,  \textbf{X}_i) = log(v_i)- 	log(\lambda_i). 
	\end{gather*}
	Using these relations, we now insert the weights into the right-hand side of equation (24) and obtain
	\begin{gather}
		\begin{split}
			&\sum_{i=1}^{n}w_ilog(\frac{w_i}{v_i})+\gamma^{'}\sum_{i=1}^{n}w_ig(\textbf{T}_i,\textbf{X}_i)+(\alpha-1)(\sum_{i=1}^{n}w_i-1)	 \\
			&= \frac{1}{\lambda}(\sum_{i=1}^{n}\lambda_i[log(\lambda_i)-log(v_i)+log(\lambda)]+\sum_{i=1}^{n}\lambda_i[log(v_i)-log(\lambda_i)])\\	
			&= \frac{log(\lambda)}{\lambda}(\sum_{i=1}^{n}\lambda_i)\\
			&= log(\lambda)=log(\sum_{i=1}^{n}v_iexp(-\gamma^{'}g(\textbf{T}_i,\textbf{X}_i))).	
		\end{split}
	\end{gather}
	This final line of Equation (26) is no longer a function of the weights. It can therefore be solved by convex algorithms. Finally, because we set $v_i$ to $\frac{1}{n}$ throughout, we can obtain the objective function presented in Section 3.2.
	
	\subsection*{A.3.2\enspace Large Sample Properties}
	\subsubsection*{A.3.1\enspace Proof of Theorem 1}
	We first show that the conclusion of Theorem(\romannumeral1). \\
	
	Since $\hat{\theta}$ (as a estimator of $\theta^*$) is a unique minimizer of $\frac{1}{n}\sum_{i=1}^{n}\hat{w_i}(Y_i-s(\textbf{T}_i;\theta))^2$(regarding $\mathbb{E}[w(Y-s(\textbf{T};\theta))^2]$, according to the theory of M-estimation (\citet{van2000asymptotic}, Theorem 5.7), if%(van der Vaart, 2000, Theorem 5.7), 
	\begin{gather*}
		\text{sup}_{\theta\in \Theta} \mid \frac{1}{n}\sum_{i=1}^{n}\hat{w_i}(Y_i-s(\textbf{T}_i;\theta))^2-\mathbb{E}[w(Y-s(\textbf{T};\theta))^2]) \mid \to_p 0,
	\end{gather*}
	then $\hat{\theta} \to_p \theta^*$.
	Note that
	\begin{gather}
		\text{sup}_{\theta\in \Theta} \mid \frac{1}{n}\sum_{i=1}^{n}\hat{w_i}(Y_i-s(\textbf{T}_i;\theta))^2-\mathbb{E}[w(Y-s(\textbf{T};\theta))^2]) \mid  \notag \\
		\leq \text{sup}_{\theta\in \Theta} \mid \frac{1}{n}\sum_{i=1}^{n}(\hat{w_i}-w_i)(Y_i-s(\textbf{T}_i;\theta))^2 \mid \notag \\
		+\text{sup}_{\theta\in \Theta} \mid \frac{1}{n}\sum_{i=1}^{n}w_i(Y_i-s(\textbf{T}_i;\theta))^2-\mathbb{E}[w(Y-s(\textbf{T};\theta))^2]) \mid.
	\end{gather}
	We first show that $\text{sup}_{\theta\in \Theta} \mid \frac{1}{n}\sum_{i=1}^{n}(\hat{w_i}-w_i)(Y_i-s(\textbf{T}_i;\theta))^2 \mid$ is $o_p(1)$. Using the Causchy-Schwarz inequality and the fact that $\hat{w}\to^{L^2} w$ (Ai, et al.(2020), Theorem 3), we have
	\begin{gather*}
		\begin{split}
			\text{sup}_{\theta\in \Theta} \mid \frac{1}{n}\sum_{i=1}^{n}(\hat{w_i}-w_i)(Y_i-s(\textbf{T}_i;\theta))^2 \mid  & \leq \lbrace  \frac{1}{n}\sum_{i=1}^{n}(\hat{w_i}-w_i)^2 \rbrace^{1/2} \text{sup}_{\theta\in \Theta} \lbrace \frac{1}{n}\sum_{i=1}^{n}(Y_i-s(\textbf{T}_i;\theta))^2 \rbrace^{1/2} \\
			& \leq o_p(1)\lbrace \text{sup}_{\theta\in \Theta} \mathbb{E}[w(Y-s(\textbf{T};\theta))^2]+o_p(1) \rbrace^{1/2}\\
			&=o_p(1).
		\end{split}
	\end{gather*}
	Thereafter, under Assumption 6, we can conclude that $\text{sup}_{\theta\in \Theta} \mid \frac{1}{n}\sum_{i=1}^{n}w_i(Y_i-s(\textbf{T}_i;\theta))^2-\mathbb{E}[w(Y-s(\textbf{T};\theta))^2]) \mid $ is also $o_p(1)$ (\citet{newey1994large}, Lemma 2.4). Hence, we complete the proof for Theorem 1(\romannumeral1). Next, we give the proof of Theorem 1(\romannumeral2). \\
	Define
	\begin{equation*}
		\hat{\theta}^* = \text{argmin}_\theta \sum_{i=1}^{n} w_i(Y_i-s(\mathbf{T}_i;\theta))^2.
	\end{equation*}
	Assume that $\frac{1}{n} \sum_{i=1}^{n} w_i(Y_i - s(\mathbf{T}_i;\hat{\theta}^*))h(\mathbf{T}_i;\hat{\theta}^*)) = o_p(n^{-1/2})$ holds with probablility to one as $n \to \infty$
		
		By Assumption 5 and the uniform law of large number, one can get that
	\begin{equation*}
	\frac{1}{n}\sum_{i=1}^{n} w_i(Y_i-s(\textbf{T}_i;\theta))^2 \to \mathbb{E} \lbrace w(Y-s(\textbf{T};\theta))^2 \rbrace \  \text{in probability uniformly over }\ \theta,
	\end{equation*}
	which implies $\mid \mid \hat{\theta}^* -\theta^* \mid \mid \to_p 0$. Let
	\begin{equation*}
	r(\theta) = 2\mathbb{E} \lbrace w(Y-s(\textbf{T};\theta))h(\textbf{T};\theta) \rbrace,
	\end{equation*}
	which ia a differentiable function in $\theta$ and $r(\theta^*) = 0$. By mean value theorem, we have
	\begin{equation*}
	\sqrt{n}r(\hat{\theta}^*)- \bigtriangledown_\theta r(\zeta) \cdot \sqrt(n)(\hat{\theta}^* - \theta^*) =\sqrt{n}r(\theta^*) =0
	\end{equation*}
	where $\zeta$ lies on the line joining $\hat{\theta}^*$ and $\theta^*$. Since $\bigtriangledown_\theta r(\theta)$ is continuous at $\theta^*$ and $\mid \mid \hat{\theta}^* -\theta^* \mid \mid \to_p 0$, then
	\begin{equation*}
	\sqrt{n}(\hat{\theta}^* - \theta^*)  = \bigtriangledown_\theta r(\theta^*)^{-1}\cdot \sqrt{n} r(\hat{\theta}^*) +o_p(1)
	\end{equation*}
Define the empirical process
	\begin{equation*}
	G_n(\theta)= \frac{2}{\sqrt{n}} \sum_{i=1}^{n} \lbrace w_i(Y_i-s(\textbf{T}_i;\theta))h(\textbf{T}_i;\theta) - \mathbb{E} \lbrace w(Y-s(\textbf{T};\theta))h(\textbf{T};\theta)  \rbrace \rbrace.
	\end{equation*}
Then we have
	\begin{equation*}
		\begin{split}
			&	\sqrt{n}(\hat{\theta}^* - \theta^*)\\
			&=  \bigtriangledown_\theta r(\theta^*)^{-1}\cdot \lbrace  \sqrt{n} r(\hat{\theta}^*) -  \frac{2}{\sqrt{n}} \sum_{i=1}^{n} \lbrace w_i(Y_i-s(\textbf{T}_i;\hat{\theta}^*))h(\textbf{T}_i;\hat{\theta}^*)  + \frac{2}{\sqrt{n}} \sum_{i=1}^{n} \lbrace w_i(Y_i-s(\textbf{T}_i;\hat{\theta}^*))h(\textbf{T}_i;\hat{\theta}^*)  \rbrace \\
			&=  -\bigtriangledown_\theta r(\theta^*)^{-1}\cdot G_n(\hat{\theta}^*)+o_p(1)\\
			&= U^{-1} \cdot \lbrace G_n(\hat{\theta}^*)-G_n(\theta^*) +G_n(\theta^*) \rbrace +o_p(1).
		\end{split}	
	\end{equation*}
By Assumption 5, 6, Theorem 4 and 5 of Andrews(1994), we have $G_n(\hat{\theta}^*)-G_n(\theta^*) \to_p o$. Thus,
	\begin{equation*}
	\sqrt{n}(\hat{\theta}^* - \theta^*) = U^{-1} \frac{2}{\sqrt{n}} \sum_{i=1}^{n} \lbrace w_i(Y_i-s(\textbf{T}_i;\theta^*))h(\textbf{T}_i;\theta^*) \rbrace +o_p(1),
	\end{equation*}
	then we can get that the asymptotic variance of $\sqrt{n}(\hat{\theta}^* - \theta^*)$ is $V$.
		Therefore, $\sqrt{n}(\hat{\theta}^* - \theta^*) \to_d N(0,V)$. Next, we will prove $\hat{\theta} \to_p \hat{\theta}^*$.
	Since
	
	\begin{gather*}
		\text{sup}_{\theta\in \Theta} \mid \frac{1}{n}\sum_{i=1}^{n}\hat{w_i}(Y_i-s(\textbf{T}_i;\theta))^2-\frac{1}{n}\sum_{i=1}^{n}w_i(Y_i-s(\textbf{T}_i;\theta))^2) \mid  \\
		\leq \text{sup}_{\theta\in \Theta} \mid \frac{1}{n}\sum_{i=1}^{n}(\hat{w_i}-w_i)(Y_i-s(\textbf{T}_i;\theta))^2 \mid \\
		\leq \lbrace  \frac{1}{n}\sum_{i=1}^{n}(\hat{w_i}-w_i)^2 \rbrace^{1/2} \text{sup}_{\theta\in \Theta} \lbrace \frac{1}{n}\sum_{i=1}^{n}(Y_i-s(\textbf{T}_i;\theta))^2 \rbrace^{1/2} \\
	\leq o_p(1)\lbrace \text{sup}_{\theta\in \Theta} \mathbb{E}[w(Y-s(\textbf{T};\theta))^2]+o_p(1) \rbrace^{1/2} \\
	=o_p(1),
	\end{gather*}
	which implies $\hat{\theta}^* \to_p \hat{\theta}$. Then by Slutskey's Theorem, we can draw the conclusion that $\sqrt{n}(\hat{\theta} - \theta^*) \to_d N(0,V)$. Therefore, we have completed the proof of Theorem 1.

	\subsubsection*{A.2\enspace Proof of Theorem 2}
	To obtain the convergence rate of the estimate $\hat{s}(\textbf{t})=B(\textbf{t})^{'}\hat{\beta}$, we first obtain the convergence rate of $\hat{\beta}$.
	\noindent
	
	Note that
	\begin{gather}
		\begin{split}
			\hat{\beta}-\beta^* &= (Z_n^{'}Z_n)^{-1}Z_n^{'}\hat{W}Y-\beta^* \\
			&= (Z_n^{'}Z_n)^{-1}Z_n^{'}(\hat{W}-W)Y+(Z_n^{'}Z_n)^{-1}Z_n^{'}(WY-E(WY\mid \textbf{T})) \\
			&+(Z_n^{'}Z_n)^{-1}Z_n^{'}(E(WY\mid \textbf{T})-Z_n\beta^*)\\
			&\equiv B_1+B_2+B_3.
		\end{split}
	\end{gather}
	First, we compute the convergence order of $B_1$. Let $H = (\hat{W}-W)Y$ and $\hat{\Lambda} = \frac{1}{n} Z_n^{'}Z_n$; then,
	\begin{gather}
		\begin{split}
			\mid\mid B_1 \mid\mid^2 &= \mid\mid(n\hat{\Lambda})^{-1}Z_n^{'}H\mid\mid^2 \\
			&= n^{-2}tr(H^{'}Z_n(\hat{\Lambda})^{-2}Z_n^{'}H) \\
			&= n^{-2}tr((\hat{\Lambda})^{-1}Z_n^{'}HH^{'}Z_n(\hat{\Lambda})^{-1})\\
			&=  n^{-2}tr((\hat{\Lambda})^{-1/2}Z_n^{'}HH^{'}Z_n(\hat{\Lambda})^{-1/2}(\hat{\Lambda})^{-1})\\
			&\leq \lambda_{max}((\hat{\Lambda})^{-1})n^{-2}tr((\hat{\Lambda})^{-1/2}Z_n^{'}HH^{'}Z_n(\hat{\Lambda})^{-1/2}) \\
			&= \lambda_{max}((\hat{\Lambda})^{-1})n^{-2}tr(HH^{'}Z_n(\hat{\Lambda})^{-1}Z_n^{'})\\
			&=  \lambda_{max}((\hat{\Lambda})^{-1})n^{-1}tr(HH^{'}Z_n(Z_n^{'}Z_n)^{-1}Z_n^{'}) \\
			& \leq \lambda_{max}((\hat{\Lambda})^{-1})n^{-1} \mid\mid H \mid\mid^2 \\
			&=  \lambda_{max}((\hat{\Lambda})^{-1})n^{-1} Y^{'}(\hat{W}-W)^{'}(\hat{W}-W)Y \\
			&= \lambda_{max}((\hat{\Lambda})^{-1}) sup_{(\textbf{t},\textbf{x})}\mid \hat{W}-W \mid^2 \frac{1}{n}Y^{'}Y \\
			&\leq O_p(1) O_p(\frac{1}{n})O_p(1) \\
			&= O_p(\frac{1}{n}),
		\end{split}
	\end{gather}
	where the first inequality follows from $tr(AB) \leq \lambda_{max}(B)tr(A)$ for any symmetric matrix $B$ and positive semidefinite matrix $A$, the second inequality follows from the fact that $Z_n(Z_n^{'}Z_n)^{-1}Z_n^{'}$ is a projection matrix with a maximum eigenvalue of 1, and the last inequality follows from the fact that $ \mid \lambda_{max}((\hat{\Lambda})^{-1})  \mid =O_p(1)$, Ai et al., 2020, Lemma 3.1, Corollary 3.3, and $\frac{1}{n}Y^{'}Y =O_p(1)$. 
	
	\noindent
	Thereafter, we compute the convergence order of $B_2$. Let $E= WY-E(WY\mid \textbf{T})$, then
	\begin{gather}
		\begin{split}
			\mid\mid B_2 \mid\mid^2 &= \mid\mid (n\hat{\Lambda})^{-1}Z_n^{'}E \mid\mid^2 \\
			&=  n^{-2}tr(E^{'}Z_n(\hat{\Lambda})^{-2}Z_n^{'}E) \\
			&= n^{-2}tr((\hat{\Lambda})^{-1}Z_n^{'}EE^{'}Z_n(\hat{\Lambda})^{-1}) \\
			&= n^{-2}tr(Z_n^{'}EE^{'}Z_n(\hat{\Lambda})^{-2}) \\
			&\leq \lambda_{max}((\hat{\Lambda})^{-2})n^{-2} tr(Z_n^{'}EE^{'}Z_n) \\
			&= \lambda_{max}((\hat{\Lambda})^{-2})n^{-2} \mid\mid E^{'}Z_n \mid\mid^2 \\
			&= O_p(\frac{Q}{n}),
		\end{split}
	\end{gather}
	where the last equality follows the fact that $ \mid \lambda_{max}((\hat{\Lambda})^{-1})  \mid =O_p(1)$ and $n^{-2} \mid\mid E^{'}Z_n \mid\mid^2  = O_p(\frac{Q}{n})$ by Markov's inequality.
	
	\noindent
	Finally, we compute the convergence order of $B_3$. Let $M = E(WY\mid \textbf{T})-Z_n\beta^*$, then
	\begin{gather}
		\begin{split}
			\mid\mid B_3 \mid\mid^2 &=  \mid\mid (n\hat{\Lambda})^{-1}Z_n^{'}M \mid\mid^2 \\
			&= n^{-2}tr(M^{'}Z_n(\hat{\Lambda})^{-2}Z_n^{'}M) \\
			&= n^{-2}tr((\hat{\Lambda})^{-1}Z_n^{'}MM^{'}Z_n(\hat{\Lambda})^{-1}) \\
			&= n^{-2}tr((\hat{\Lambda})^{-1/2}Z_n^{'}MM^{'}Z_n(\hat{\Lambda})^{-1/2}(\hat{\Lambda})^{-1}) \\
			&\leq \lambda_{max}((\hat{\Lambda})^{-1})n^{-2}tr((\hat{\Lambda})^{-1/2}Z_n^{'}MM^{'}Z_n(\hat{\Lambda})^{-1/2}) \\
			&= \lambda_{max}((\hat{\Lambda})^{-1})n^{-2}tr(Z_n^{'}MM^{'}Z_n(\hat{\Lambda})^{-1}) \\
			&= \lambda_{max}((\hat{\Lambda})^{-1})n^{-1}tr(MM^{'}Z_n(Z_n^{'}Z_n)^{-1}Z_n^{'})\\
			&=  \lambda_{max}((\hat{\Lambda})^{-1})n^{-1}\sum_{i=1}^{n}\lbrace E(w_iY_i\mid \textbf{T}_i)-B(\textbf{T}_i^{'})\beta^*\rbrace^2.
		\end{split}
	\end{gather}
	Since
	\begin{gather*}
		n^{-1}\sum_{i=1}^{n}\lbrace E(w_iY_i\mid \textbf{T}_i)-B(\textbf{T}_i^{'})\beta^*\rbrace^2 \leq \text{sup}_{\textbf{t}\in \mathcal{T}}\mid E(wY\mid \textbf{T}=\textbf{t}) - B(\textbf{t})^{'}\beta^* \mid^2 = O_p(Q^{-2\tilde{\alpha}}),
	\end{gather*}
	we can deduce that $\mid\mid B_3 \mid\mid^2 \leq O_p(Q^{-2\tilde{\alpha}})$.
	
	\noindent
	By combining Equations (29), (30) and (31), we have
	$\mid\mid \hat{\beta} -\beta^* \mid\mid =O_p(\sqrt{\frac{Q}{n}}+Q^{-\alpha})$. \\
	
	\noindent
	
	Next, the proof of the convergence of $\hat{s}(\mathbf{t})$ is provided.	
	Let $\Lambda = E[B(\textbf{t})^{'}B(\textbf{t})]$, then we have
	\begin{gather}
		\begin{split}
			\int_\mathcal{T} \mid \hat{s}(\textbf{t})-s(\textbf{t}) \mid^2 dF_\textbf{T}(\textbf{t})&= 	\int_\mathcal{T} \mid B(\textbf{t})^{'}\hat{\beta} - B(\textbf{t})^{'}\beta^*+B(\textbf{t})^{'}\beta^*-s(\textbf{t}) \mid^2 dF_\textbf{T}(\textbf{t}) \\
			&\leq 2(\hat{\beta}-\beta^*)^{'}\int_\mathcal{T}[B(\textbf{t})^{'}B(\textbf{t})]dF_\textbf{T}(\textbf{t})(\hat{\beta}-\beta^*)+2\int_\mathcal{T}\mid B(\textbf{t})^{'}\beta^*-s(\textbf{t}) \mid^2 dF_\textbf{T}(\textbf{t})\\
			&\leq 2\mid\mid \hat{\beta}-\beta^* \mid\mid^2 \lambda_{max}(\Lambda)+2\int_\mathcal{T}\mid B(\textbf{t})^{'}\beta^*-s(\textbf{t}) \mid^2 dF_\textbf{T}(\textbf{t})\\
			&=O_p( \frac{Q}{n} + Q^{-2\tilde{\alpha}})
		\end{split}
	\end{gather}
	
	and
	\begin{gather}
		\begin{split}
			\text{sup}_{\textbf{t}\in \mathcal{T}} \mid \hat{s}(\textbf{t})-s(\textbf{t}) \mid &= 	\text{sup}_{\textbf{t}\in \mathcal{T}} \mid B(\textbf{t})^{'}\hat{\beta} - B(\textbf{t})^{'}\beta^*+B(\textbf{t})^{'}\beta^*-s(\textbf{t}) \mid \\
			&\leq \text{sup}_{\textbf{t}\in \mathcal{T}} \mid\mid B(\textbf{t}^{'}) \mid\mid \cdot \mid\mid \hat{\beta}-\beta^* \mid\mid + \text{sup}_{\textbf{t}\in \mathcal{T}}  \mid B(\textbf{t})^{'}\beta^*-s(\textbf{t}) \mid \\
			&\leq O_p(\sqrt{\frac{Q}{n}}+Q^{-\tilde{\alpha}} \rbrace)+O_p(Q^{-\tilde{\alpha}}) \\
			&= O_p(\sqrt{\frac{Q}{n}}+Q^{-\tilde{\alpha}} \rbrace),
		\end{split}
	\end{gather}
	Therefore, the proof of Theorem 2 is complete.
	
\end{document}